\documentclass[reprint,superscriptaddress,showpacs,amsmath,amssymb,aps,prl]{revtex4-1}
\usepackage{graphicx}
\usepackage{dcolumn}
\usepackage{bm}
\usepackage{hyperref}
\hypersetup{colorlinks=true, citecolor=blue, urlcolor=blue, linkcolor=blue}
\usepackage{amssymb}
\usepackage{color}
\usepackage{graphicx}
\usepackage{amsmath}
\usepackage{mathrsfs}
\usepackage{times}
\usepackage{subeqnarray}
\usepackage{cases}
\usepackage{bm}
\usepackage{diagbox}
\usepackage{booktabs}
\usepackage[table,xcdraw]{xcolor}
\usepackage{colortbl}
\usepackage{epstopdf}
\definecolor{tabcolor}{rgb}{.105,.410,.113}
\begin{document}
	
\title{Emergence of cooperation under punishment: A reinforcement learning perspective}
	
	\author{Chenyang Zhao}
	\affiliation{School of Physics and Information Technology, Shaanxi Normal University, Xi'an 710061, P. R. China}
	\author{Guozhong Zheng}
	\affiliation{School of Physics and Information Technology, Shaanxi Normal University, Xi'an 710061, P. R. China}
	\author{Chun Zhang}
	\affiliation{College of Science, Xi'an Shiyou University, Xi'an 710065, P. R. China}
        \author{Jiqiang Zhang}
	\affiliation{School of Physics and Electronic-Electrical Engineering, Ningxia University, Yinchuan 750021, P. R. China}
	\author{Li Chen}
	\email[Email address: ]{chenl@snnu.edu.cn}
	\affiliation{School of Physics and Information Technology, Shaanxi Normal University, Xi'an 710061, P. R. China}
	
\begin{abstract}
Punishment is a common tactic to sustain cooperation and has been extensively studied for a long time. 
While most of previous game-theoretic work adopt the imitation learning where players imitate the strategies who are better off, the learning logic in the real world is often much more complex. In this work, we turn to the reinforcement learning paradigm, where individuals make their decisions based upon their past experience and long-term returns. 
Specifically, we investigate the Prisoners' dilemma game with Q-learning algorithm, and cooperators probabilistically pose punishment on defectors in their neighborhood. Interestingly, we find that punishment could lead to either continuous or discontinuous cooperation phase transitions, and the nucleation process of cooperation clusters is reminiscent of the liquid-gas transition.
The uncovered first-order phase transition indicates that great care needs to be taken when implementing the punishment compared to the continuous scenario.
\end{abstract}
	
\date{\today }
\maketitle
\section{1. Introduction}\label{sec:introduction}
	
Cooperation is ubiquitous and stands out as a pivotal phenomenon among many other activities in human society~\cite{Nielsen1985cooperation,Colman1995Theory, Nowak2006Evolution}. As seen, cooperation permeates all facets of human society, and is instrumental in fostering the stability and sustainability of our civilization. However, according to ``survival of the fittest" in Darwinism, there is no place for cooperation, as the cooperators help others at a certain cost, placing themselves at a disadvantageous position. Therefore, decoding the mechanism of cooperation has been listed as one of twenty-five grand scientific questions in this century~\cite{Pennisi2005How}, which has become a highly interdisciplinary field.
	
With the help of evolutionary game theory, tremendous progress has been made and many potential mechanisms are uncovered~\cite{Nowak2006Five}. These include kin selection~\cite{Dawkins2006selfish, Wilson1975Sociobiology}, direct reciprocity~\cite{Nowak2008Repeated}, indirect reciprocity~\cite{Perc2013Interdependent, Nowak1998indirect, Ohtsuki2006indirect}, network reciprocity~\cite{Nowak1992spatial, Szabo1998Evolutionary, Wang2013Interdependent}, and group selection~\cite{Smith1964Group, Charlesworth2000Levels}. In addition, factors such as repeated interactions~\cite{Antonioni2017Coevolution, Nowak2004Emergence}, memory effects~\cite{Wang2006Memory, Qin2008Effect, Liu2010Memory, Zhang2023emergence}, social diversity~\cite{Perc2008Social, Santos2008Social,Liang2021Social}, dynamical reciprocity~\cite{Liang2022dynamical}, rewards~\cite{Sigmund2001Reward}, 
and behavioral multimodality~\cite{Ma2023emergence} may also contribute to emergence of cooperation. Altogether, these endeavors provide insights into the understanding of the multifaceted nature of cooperation.
		
In particular, previous works based on game-theoretic models~\cite{Takesue2018Evolutionary,Szolnoki2013Effectiveness,Lucas2021Symbiotic} and experiments\cite{Herrmann2008Antisocial,Diekmann2015Punitive,Fehr2002Altruistic,Henrich2006Costly,Herrmann2008Antisocial,Sigmund2007Punish} have shown that punishment is an effective way to promote cooperation prevalence across human societies. Punishment acts as a form of retaliation that is typically administered in two manners. While in the peer punishment~\cite{Brandt2003Punishment,Yang2015Peer}, the costs incurred by the punishment are borne by the player who inflicts the punishment, within the pool punishment~\cite{Sigmund2010Social,Szolnoki2011Phase}, the costs are shared by those who inflict punishment. 
In recent years, significant attention has been directed towards understanding the impact of punishment on the evolution of cooperation~\cite{Perc2017Statistical,Amor2011Effects,Boyd2003evolution,Yang2015Role,Yang2017Phase}, mostly with the public goods game.
Helbing et al.~\cite{Helbing2010Punish,Helbing2010Evolutionary,Helbing2010Defector} discussed three scenarios where punishment is imposed by cooperators, defectors, or simultaneously by both and revealed complex dependence on the punishment.
In a recent study,  the punishment coevolves with the investment, which is determined by the particle swarm optimization~\cite{LV2022Particle}. 
Besides, punishment and reward are also posed simultaneously to act synergistically to promote cooperation~\cite{Szolnoki2013Correlation,Sasaki2015Voluntary}.  These works demonstrate the crucial role of punishment in the emergence of cooperation.
Note that, most of these studies are conducted in the imitation learning framework~\cite{Roca2009Evolutionary, Szolnoki2009Topology}, where players imitate the strategies of those who have higher payoffs. Essentially, the imitation learning is a simple version of social learning~\cite{Bandura1977social}, where individuals learn from others based upon their observations and utilities in their socio-economic lives.

As a different paradigm~\cite{Kaelbling1996Reinforcement}, reinforcement learning (RL) enables the player to make one's decision based on learnt policy through the interaction with the environment. As a classical RL algorithm -- Q-learning~\cite{Watkins1992Q-learning}, players score different actions for given states, they take actions probabilistically based on these scores. These scores are stored in a repository, referred to as a Q-table, and this table is revised continuously by learning, aiming to maximize the expected payoffs. Thus, the RL paradigm is an endogenously self-reflective learning rather than resorting to other players, making the RL a fundamentally distinct paradigm  compared to the social learning. Even so, only recently, there are some works starting to integrate the RL and the evolutionary game theory framework to study many puzzles in human behaviors, including cooperation~\cite{Zhang2020Understanding, Zhao2022Reinforcement, Wang2022Levy, Wang2023Synergistic, Ding2023emergence}, trust~\cite{Zheng2023decoding}, the resource allocation~\cite{Zhang2019reinforcement, Zheng2023optimal} and so on. Although new insights are obtained for the emergence of cooperation for either two players~\cite{Ding2023emergence} or in many-player populations~\cite{Zhang2020Understanding} or with factors such as uncertainties~\cite{Wang2022Levy} or adaptive reward~\cite{Wang2023Synergistic}, there is a lack of understanding in the practice of punishment on cooperation from the RL perspective. In particular, we are interested in the following questions: \emph{how does punishment influence cooperation in this new paradigm, and is there any new physics therein?}

In this work, we investigate the impact of punishment in Prisoners' dilemma game in the paradigm of reinforcement learning. Specifically, we adopt the Q-learning algorithm, where the action of cooperation and defection for each player is guided by a Q-table. Punishment is probabilistically imposed by cooperators in the form of peer punishment with a given cost to those who administer punishment. Surprisingly, we reveal that the practice of punishment is shown to have rich cooperation phase transitions, which could be either continuous or discontinuous as the punishment intensity is varied, depending on the cost, and vice versa. Physically, there is a nucleation process during the formation of cooperation clusters in the discontinuous transition. Mechanistic analysis based on the evolution of Q-table shows that there are crossovers in the cooperation preference that are responsible for the abrupt transitions.

\section{2. Model}\label{sec:model}
	
We start by introducing the Prisoner's Dilemma (PD) game, where two players are engaged and they must decide simultaneously and independently whether to cooperate (C) or to defect (D). The payoffs are as follows: both receive a reward $R$ for mutual cooperation, or a punishment payoff $P$ for mutual defection, and if one cooperates while the other defects, the defector receives a temptation payoff of $T$, while the cooperator receives a sucker's payoff of $S$. We adopt the strong version of PD~\cite{Robert1981Evolution}, these payoffs satisfy the order of $T\textgreater R \textgreater P \textgreater S$ and the payoff matrix is configured as follows:
\begin{equation}
		\begin{pmatrix}
			\Pi _{CC}  & \Pi _{CD}\\
			\Pi _{DC} &\Pi _{DD}
		\end{pmatrix}=\begin{pmatrix}
			R & S\\
			T & P
		\end{pmatrix}=\begin{pmatrix}
			1 & -b\\
			1+b & 0
		\end{pmatrix},
		\label{eq:matrix}
\end{equation}
where $b\in(0,1)$, controlling the strength of the dilemma. 
	
Specifically, the players adopt the Q-learning algorithm~\cite{Watkins1992Q-learning,Sutton2018reinforcement}, where each has a Q-table in hand that is policy collection guiding one's decision-making, see Table~\ref{tab:Qtable}. The idea of Q-table is to score each action $a\in\mathcal A$ within a given state $s\in \mathcal S$ in the form of its item $Q_{s,a}$. Here, $\mathcal A$ =$\left \{C, D\right \}$ denotes as the action set, and $\mathcal S$ = $\left \{ s_{0} ,s_{1},...,s_{5} \right \}$ as the state set, where we tally the number of cooperators in the focal player's neighbourhood including itself, i.e. $s_0=0$, $s_1=1$, ..., $s_5=5$. The score $Q_{s,a}$ is the value function to estimate the value of an action $a$ given the state $s$. $Q_{s,a}>Q_{s,\hat{a}}$ means a higher value of the corresponding action $a$ for the given state $s$ than the other $\hat{a}$, and the action $a$ is thus more preferred. In Q-learning, every player aims to optimize its Q-table to maximize its accumulated reward.

\begin{table}[]
\begin{tabular}{c|cc}
\arrayrulecolor{tabcolor}\toprule [1.4pt]
\hline
\diagbox{State}{Action}& $a_{1}$ = C & $a_{2}$ = D \\
\midrule [0.5pt]
\hline
0 $(s_{0})$ & $Q_{s_{0},a_{1}}$ & $Q_{s_{0},a_{2}}$  \\
1 $(s_{1})$ & $Q_{s_{1},a_{1}}$ & $Q_{s_{1},a_{2}}$ \\
2 $(s_{2})$ & $Q_{s_{2},a_{1}}$ & $Q_{s_{2},a_{2}}$ \\
3 $(s_{3})$ & $Q_{s_{3},a_{1}}$ & $Q_{s_{3},a_{2}}$\\   
4 $(s_{4})$ & $Q_{s_{4},a_{1}}$ & $Q_{s_{4},a_{2}}$  \\  
5 $(s_{5})$ & $Q_{s_{5},a_{1}}$ & $Q_{s_{5},a_{2}}$  \\  
\hline
\bottomrule[1.4pt]
\end{tabular}
\caption{Q-table for every player. The state $s_{1,...,5}$ corresponds to the number of cooperators in its neighborhood, i.e. the four nearest neighbors plus the player itself, and there are two choices (C or D) for the action $a_{1,2}$.}
\label{tab:Qtable}
\end{table}
	
We place $N$ players on a $L\times L$ square lattice ($N=L\times L$) with a periodic boundary condition. 
They play the PD game with their four nearest neighbors, and the evolution follows a synchronous updating scheme.  
Initially, each player is endowed with a random strategy from the set $\mathcal A$, and a Q-table with all items $Q_{s,a}$ being chosen independently and randomly from $0$ to $1$. 
At the beginning of each round $t$, those who choose cooperation might not be satisfied with the outcome are assumed to impose punishment with a probability $\rho$ against all defectors in their neighborhood. The punished defectors are then subjected to a punishment intensity in terms of penalty $p$, and the cooperator who administers the punishment incurs a cost of $c$ for each penalty. Notice that, in this case defectors might be subject to multiple punishment simultaneously, and cooperators who punish could also bear multiple costs depending on the number of conducted punishments. 

Next to that, the players make their moves and revise their Q-tables. With an exploration probability $\epsilon$, each player randomly selects an action from the set $\mathcal A$ to account for trail-and-error exploration. Otherwise, players strictly follow the guidance of their Q-tables, they select the action $a$ with the corresponding $Q_{s,a}$ being larger than the other in the given row.
The decision-making phase is completed. Players then proceed to update their Q-tables as follows:
\begin{equation}
		\label{eq:Q_updating}
		Q_{s,a}(t+1) = (1-\alpha)Q_{s,a}(t)+\alpha [\Pi_t+\gamma \max_{a'}Q_{s',a'}(t)].
\end{equation}
Here, $s$ and $a$ denote the player's state and action in the round $t$, while $s'$ and $a'$ represent the new state and action at $t+1$. 
The learning rate $\alpha\in(0,1)$ determines how much of Q-value is renewed; $\alpha\rightarrow0$ means that nearly nothing is absorbed into the Q-value, while $\alpha\rightarrow 1$ means that the old experience stored in $Q_{s,a}(t)$ is completely replaced.
The discount factor $\gamma\in(0,1)$ weighs the importance of future rewards, a large value of $\gamma$ implies the player appreciates the return in the future, which can be interpreted as having a long-term vision. 
$\Pi_t$ represents the reward obtained in the round $t$, and there are two kinds of rewards accordingly: 
\begin{equation}
	\left\{
		\begin{array}{l}
			\Pi_{C} = \Pi, \\
			\Pi_{D} = \Pi - p \times n_{_{CP}}, \\
		\end{array}
	\right.
	\label{reward}
\end{equation}
where $\Pi_{C,D}$ are respectively the payoff for cooperators and defectors, and $\Pi$ is the payoff following the game matrix Eq. (\ref{eq:matrix}). Here, $p$ is the punishment intensity and $n_{_{CP}}$ denotes the number of cooperators who chose to punish in the neighborhood of the focal defector. But for those cooperators who conduct punishments, the borne cost has to be subtracted, i.e. $\Pi_{C} = \Pi - c\times n_{D}$. $c$ is the cost for conducting a single punishment and $n_D$ is the number of defectors around a focal cooperator who chooses to inflict punishment. 
While $Q_{s,a}(t)$ and $\Pi_t$ respectively represent the past experience and the reward obtained at current step in Eq.~(\ref{eq:Q_updating}),  ${\max_{a'}}Q_{s',a'}(t)$ is the maximal Q-value one can expect in the new round.

In our study, we fix the learning rate $\alpha = 0.1$, a discount factor $\gamma = 0.9$, and the exploration rate $\epsilon = 0.01$ for the Q learning algorithm. For the PD game, $b=0.1$, and the system size $L=100$ if not stated otherwise. For each simulation, a transient of at least $10^7$ steps is used, and then we average over another 500 steps for reliable results.
\begin{figure}[tbp]
		\centering
		\includegraphics[width=0.9\linewidth]{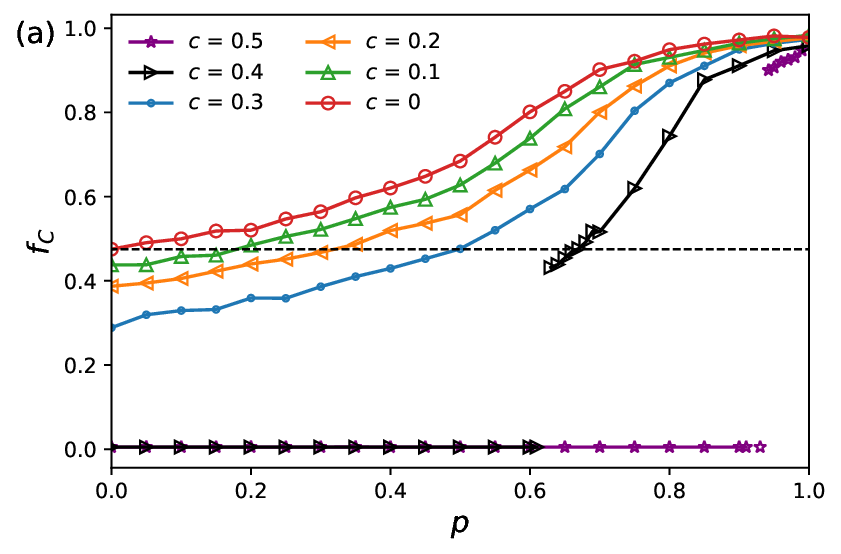}
		\includegraphics[width=0.9\linewidth]{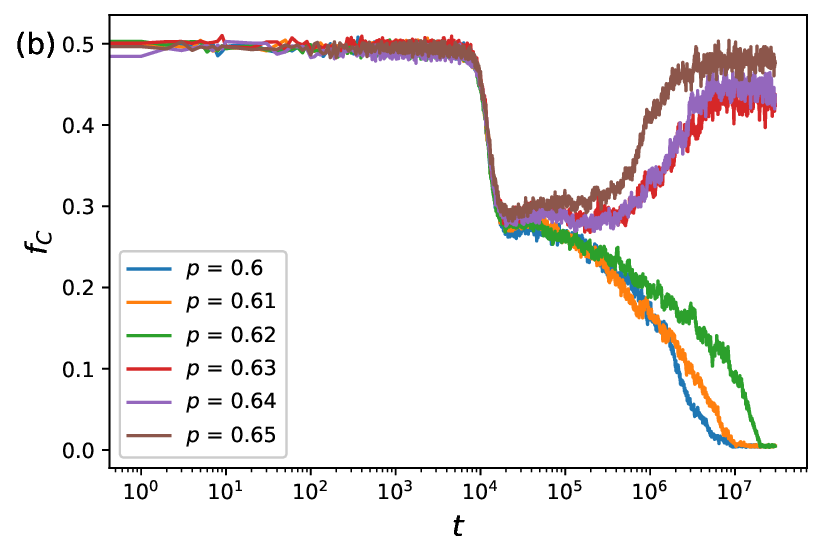} 
		\caption{
			\textbf{The impact of punishment intensity on the evolution of cooperation.}
			 (a) The cooperation level $f_C$ versus the punishment intensity $p$ for different costs $c$. The black dashed line is the cooperation level for the benchmark scenario where no punishment is conducted (i.e. $p=c=0$).
			 (b) Time series of $f_C$ for some punishment intensities by fixing $c=0.4$, corresponding to a discontinuous transition in (a).
			Other parameters: $b=0.1$, $\rho=0.2$.}
		\label{fig:p}
\end{figure}

\begin{figure}[bpth]
	\centering
		\includegraphics[width=0.9\linewidth]{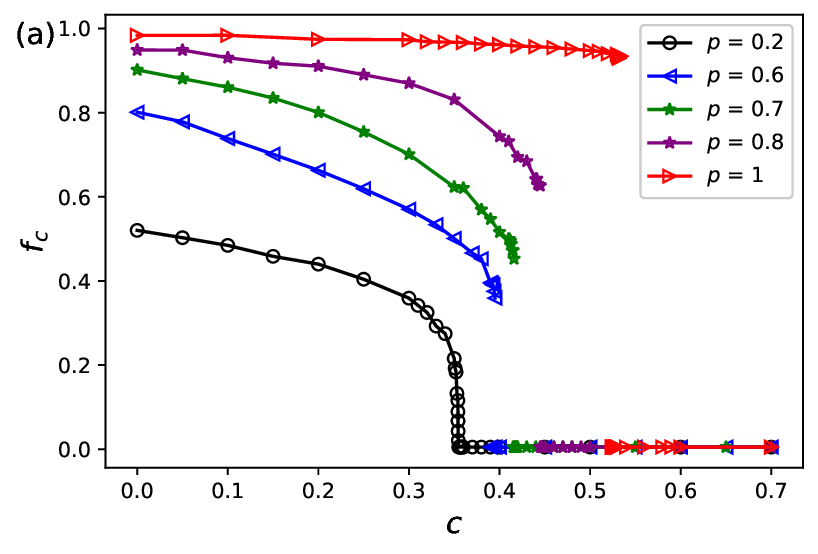} 
		\includegraphics[width=0.9\linewidth]{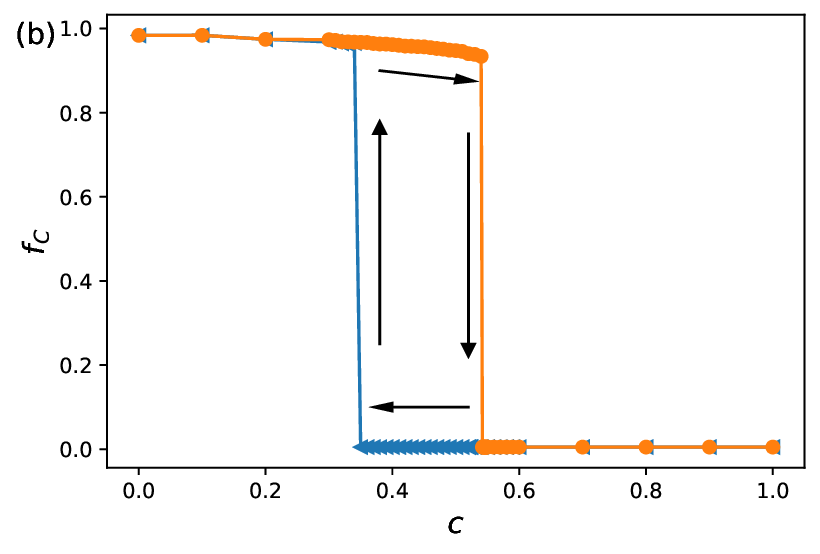}
		\caption{
		\textbf{The impact of the costs and hysteresis.}
			(a) The cooperation level $f_C$ versus the cost $c$ for different punishment intensities $p$. 
			(b) The cooperation hysteresis $f_C$ when the cost $c$ is varied in different directions for the punishment intensity $p=1$. 
			Parameters: $\emph{b}$ = 0.1, $\rho = 0.2$.
			}
		\label{fig:cost}
\end{figure}

\begin{figure*}[tbp]
		\centering
		\includegraphics[width=1.0in]{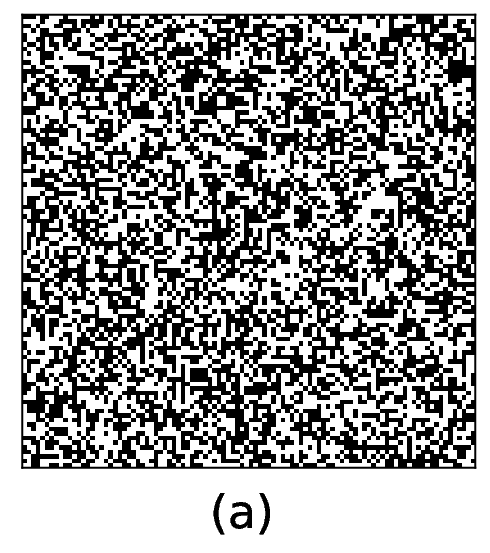}
		\includegraphics[width=1.0in]{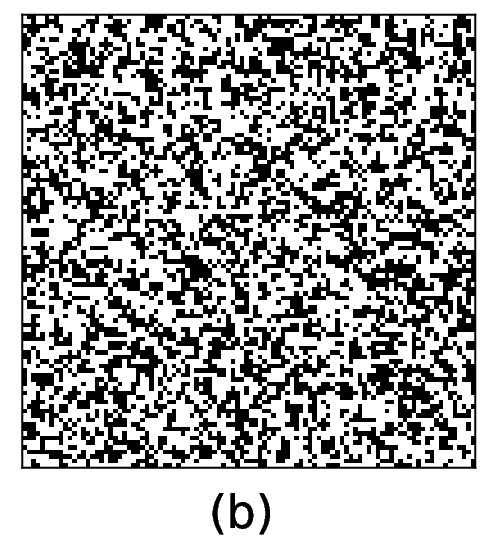}
		\includegraphics[width=1.0in]{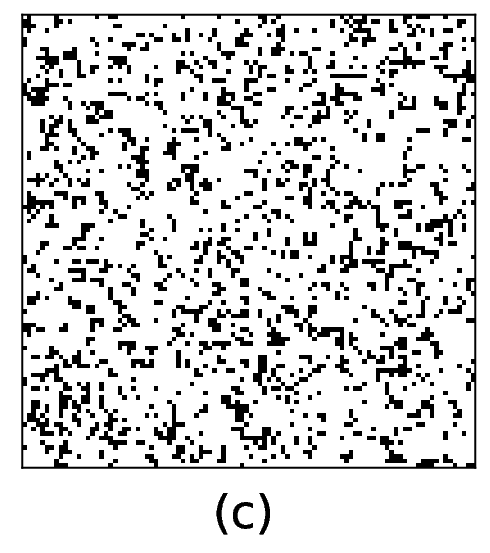}
		\includegraphics[width=1.0in]{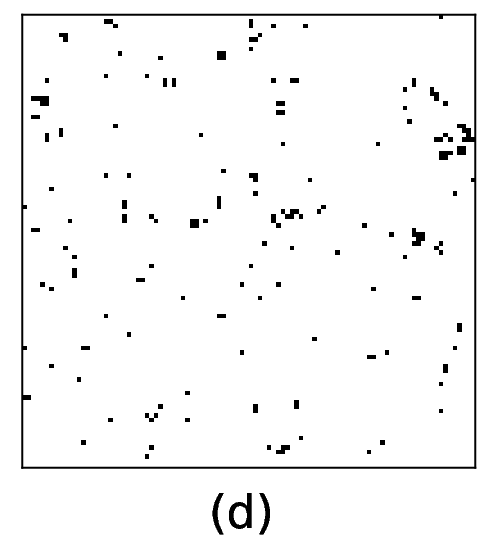}\\
		\includegraphics[width=1.0in]{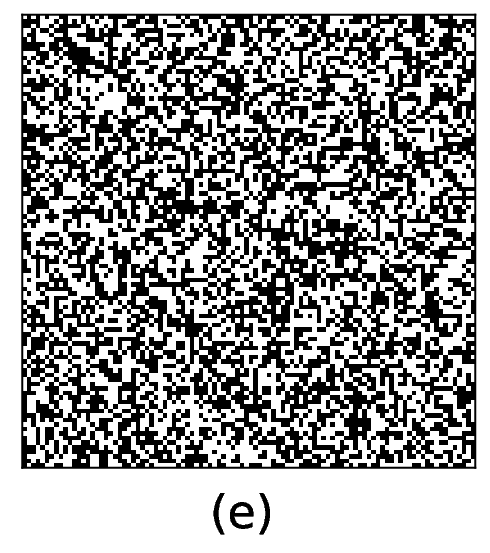}
		\includegraphics[width=1.0in]{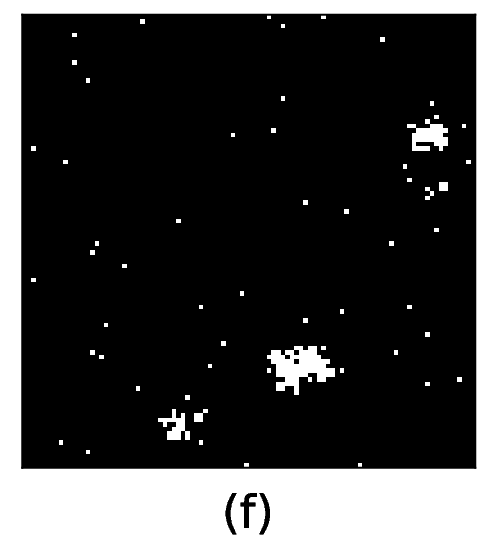}
		\includegraphics[width=1.0in]{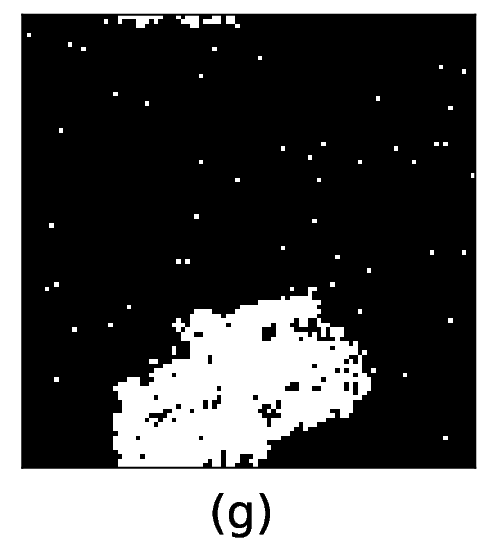}
		\includegraphics[width=1.0in]{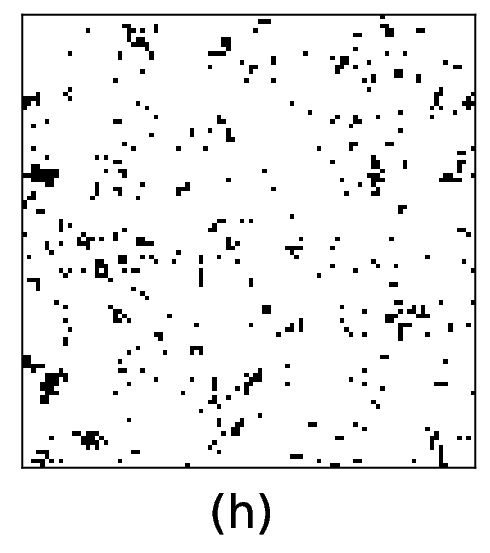}\\
		\includegraphics[width=1.0in]{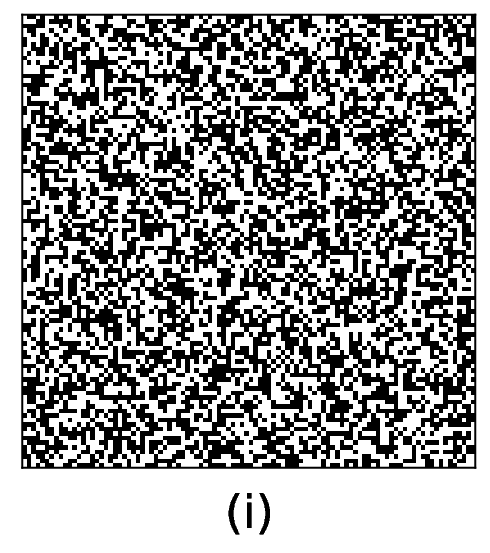}
		\includegraphics[width=1.0in]{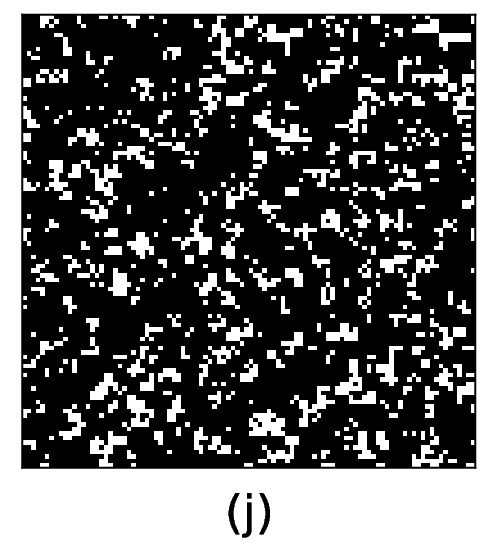}
		\includegraphics[width=1.0in]{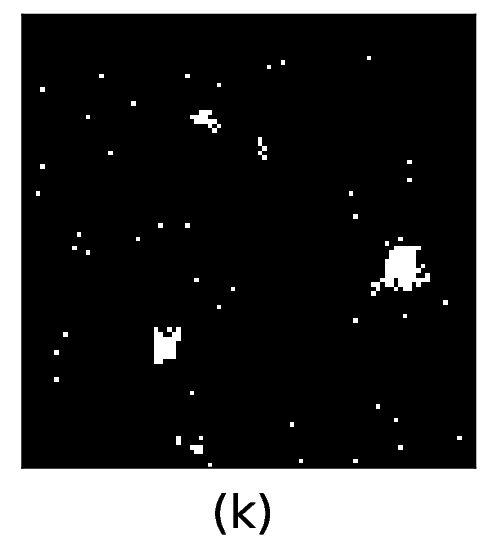}
		\includegraphics[width=1.0in]{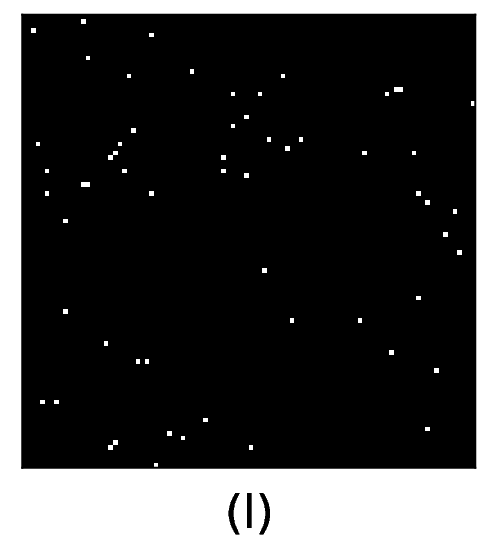}
		\caption{
		\textbf{Typical snapshots for cooperation patterns.} Cooperators and defectors are color-coded in white and black.
		(a-d)	are for $c=0.2$ outside of the bistable region, where cooperation thrives gradually. (e-h) and (i-l) are for $c=0.525$ locating in the bistable region, where the nucleation of cooperators are both seen, the cooperation thrives in panels (e-h) but vanishes in panels (i-l) . Other parameter: $p=1$.
		}
		\label{fig:snapshot} 
\end{figure*}
\section{3. Results}\label{sec:results}
	
We first illustrate the impact of punishment intensity on the level of cooperation $f_C$ with different fixed costs using PD game, shown in Fig.~\ref{fig:p}. 
In the absence of punishment (i.e. $c=0$, and $p=0$), the cooperation level is around $f_C\approx47\%$ (the horizontal dashed line), meaning that a certain level of cooperation already emerges within the RL paradigm, where the players learn to cooperate by self-reflection. This observation is in line with previous works within two-player setup~\cite{Ding2023emergence}, where cooperation emerges for the given learning parameters, and can be taken as the benchmark scenario since no any punishment is engaged. 
When a given punishment is posed $p>0$,  the case with a lower cost $c$ yields a higher cooperation level as expected, and an escalating cost leads to a decrease in $f_C$, see Fig.~\ref{fig:p}(a).
This is because opting to punish incurs a cost to the cooperators averagely, a higher cost diminishes their preference in the action of cooperation. 

Interestingly, for a given cost, the dependence of cooperation level on the punishment intensity shows two distinct types of phase transitions. At low costs, 
the cooperation level continuously increases with escalating punishment intensity, see Fig.~\ref{fig:p}(a). 
However, at high costs, the system undergoes an abrupt transition from defection to cooperation.
$f_C$ consistently remains in the almost full defection state $f_C\approx0$ until a bistable region is reached, where both the high cooperation state and  the almost full defection state are possible, hinging on initial conditions. 
Beyond that region, the population stays in a state of high cooperation, marking a discontinuous change. This means that, if the cost is high, low intensity of punishment is actually a bad choice where individuals are all choose not to cooperate, avoid conducting punishments. From the systematic point of view, the conduct of punishment fails because the resulting cooperation level $f_C$ is much lower than the level in the benchmark scenario. 
	
To more intuitively understand the discontinuous cooperation phase, we provide some time series around the threshold $p_c$ by fixing $c=0.4$, see Fig.~\ref{fig:p}(b). As can be seen, the plain around one-half at the initial stage are all followed by a decline in $f_C$; the cases with $p=0.6,0.61$, and $0.62$ continually decrease to zero, but in the other cases with a slighter larger value ($p=0.63,0.64, 0.65$), there is a turn-up in the tend till a  saturation at a high cooperation level is reached. This discrepancy in the evolution leads to the discontinuity of $f_C$ and this process is reminiscent of first-order phase transition in many systems~\cite{Sethna2021statistical, Cai2015avalanche}. By extrapolation, when the punishment cost is exceedingly high, the population remains in the almost full defection state only if a strong enough intensity of punishment is posed, beyond which the population enters almost full cooperation $f_C\approx1$.

We next turn to the impact of the cost, which is equally crucial for implementing the punishment. Fig.~\ref{fig:cost}(a) shows the fraction of cooperators as a function of cost for a couple of punishment intensities. We first observe a decline in $f_C$ as the cost of punishment increases, in line with our expectation. With a weak intensity (e.g. $p=0.2$), this decline shows a continuous decrease with escalating cost. With a strong intensity (e.g. $p=0.6, 0.7, 0.8,1$), a first-order phase transition is seen, where there is bistable region for the high cooperation and almost full defection states. Further increase in $c$ leads to the almost full defection where everyone learns not to be cooperators to avoid the high cost for the probabilistic punishment. Fig.~\ref{fig:cost}(b) explicitly demonstrates a hysteresis structure for the case of $p=1.0$ where the cost $c$ is continually increased from zero to unity, and then is altered in the opposite direction. The presence of hysteresis indicates the cooperation level depends on the history, which is another signature for first-order phase transitions.

\section{4. Mechanism analysis}\label{sec:Mechanism}
To understand how the cooperation is promoted by the implementation of punishment, and how the first-order phase transition emerges, we move to the mechanism analysis by examining typical spatio-temporal patterns and the evolution of Q-tables, respectively in the following two subsections.

\subsection{4.1  Spatio-temporal patterns}

To develop some intuition for the evolution of cooperation and especially for the discontinuous phase transitions, we provide some typical snapshots, shown in Fig.~\ref{fig:snapshot}. As seen, although cooperators thrive both in Fig.~\ref{fig:snapshot}(a-d) and Fig.~\ref{fig:snapshot}(e-h) in the long run, there are distinct features in evolution. While the cooperators continuously increase in Fig.~\ref{fig:snapshot}(a-d) from the very beginning, the number of cooperators in Fig.~\ref{fig:snapshot}(e-h) is reduced at the initial stage, later the cooperators form some small clusters, and then they grow and merge with each other, eventually the domain is dominated by the cooperators. This nucleation process, however, could also possibly end with the extinction of cooperators, and the system evolves into the state of almost full defection [Fig.~\ref{fig:snapshot}(i-l)], although this evolution shares exactly same parameters with Fig.~\ref{fig:snapshot}(e-h). Physically, this process is reminiscent of nucleation process in the liquid-gas transition~\cite{Sethna2021statistical}, where the destination for system evolution could be very different  in the bistable region, a signature for first-order phase transitions.
	
\begin{figure}[bpth]
		\centering
		\includegraphics[width=0.9\linewidth]{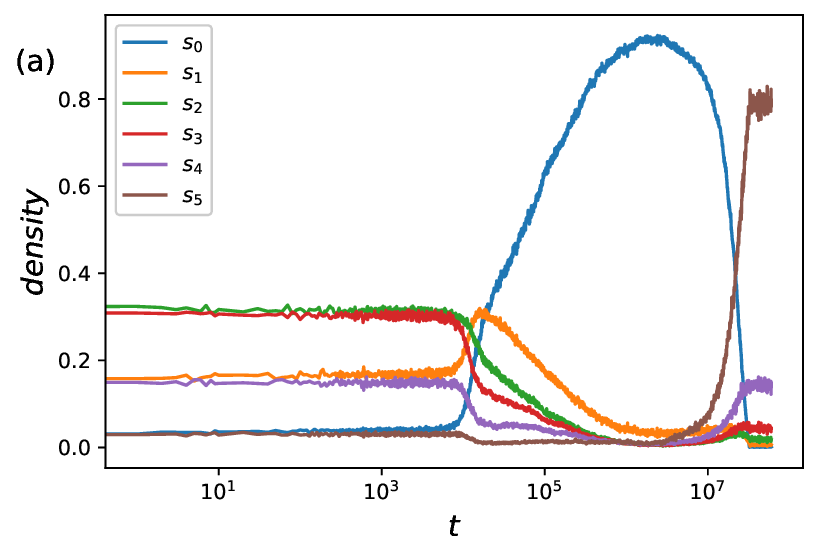}
		\includegraphics[width=0.9\linewidth]{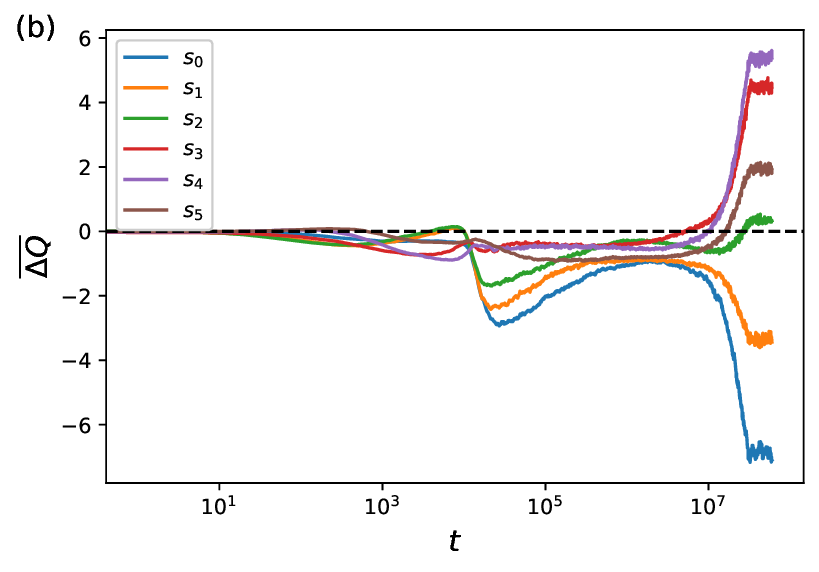} 
		\caption{
			\textbf{Evolution of Q-tables.}
			(a) The temporal evolution of all six state densities. (b) The temporal evolution of the average of Q-value difference defined as $\Delta Q=Q_C-Q_D$ for all six states. Parameters: $\emph{b}$ = 0.1, $\rho= 0.2$, $\emph{p}$ = 1, and $\emph{c}$ = 0.525.
			}
		\label{fig:Qtable}
\end{figure}

\subsection{4.2  The evolution of Q-tables}
To understand why players form cooperation clusters in Fig.~\ref{fig:snapshot}(e-h), we monitor the evolution of the corresponding Q-tables. As seen in Fig.~\ref{fig:Qtable}(a), the densities of all six states remain largely unchanged in the initial stage of evolution $t<10^4$, but then the density of $s_0$ starts to rise and becomes much higher than all others'. A longer evolution shows that the density $s_5$ overturns this trend and dominates in the end. The shown here echoes the evolution of cooperation fraction and the patterns above.  

More insights can be obtained by monitoring the cooperation preference defined as the average Q-value difference $\overline{\Delta Q_{s_j}} = \frac{1}{N} {\sum_{i=1}^{N}} (Q_{s_j,C}^{i} - Q_{s_j,D}^{i})$, where $j=0,...,5$ for different states and $i$ is the individual's label. $\overline{\Delta Q_{s_j}}>0$ means that cooperation is preferred on average within state $s_j$, otherwise defection is favored. With this in mind, let's monitor the temporal evolution of $\overline{\Delta Q_{s_j}}$ within all six states, see Fig.~\ref{fig:Qtable}(b).
 In the early stage of learning, no preference is developed as $\overline{\Delta Q}$ are consistently around $0$. Afterwards, all players become inclined to defect as all $\overline{\Delta Q_{s_j}} < 0$, which explains the rising density of $s_0$ in Fig.~\ref{fig:Qtable}(a) and thus the decline of $f_C$ in Fig.~\ref{fig:p}(b). However, the evolution in the long-term shows that there are crossovers for the preference within four states, these preferences turn to be cooperative as $\overline{\Delta Q_{s_j}} > 0$ for $j=2,3,4,5$. The other two remain negative $\overline{\Delta Q_{s_{j}}} < 0$ for $j=0,1$, which can be interpreted as a kind of ``punishment" against defectors within surroundings of a too low cooperation level. Put together, this observation shows that players learn to reciprocate cooperation surroundings with cooperation, but choose defection against defective surroundings. 
 
\begin{figure}[tpb]
		\centering
		\includegraphics[width=0.35\linewidth]{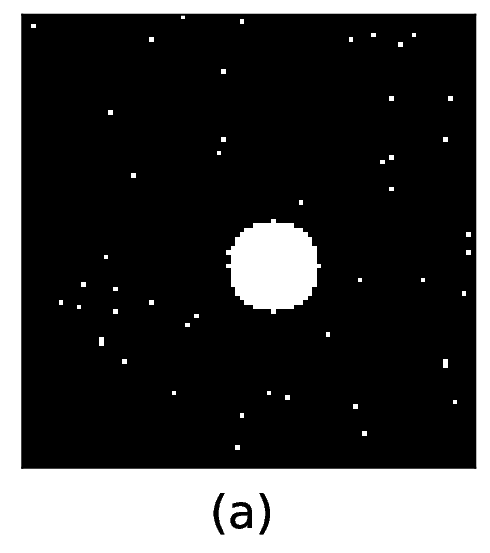}
		\includegraphics[width=0.35\linewidth]{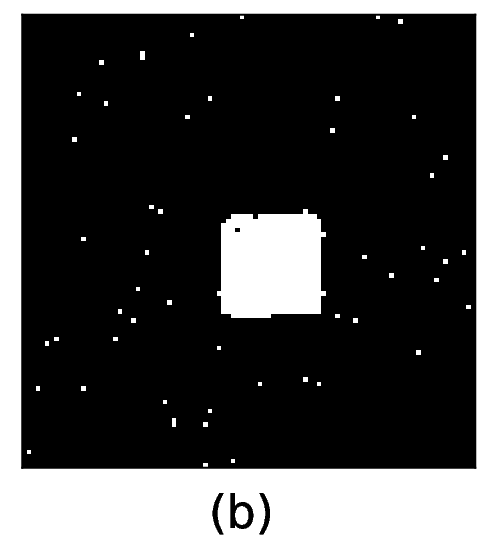}
		\includegraphics[width=0.35\linewidth]{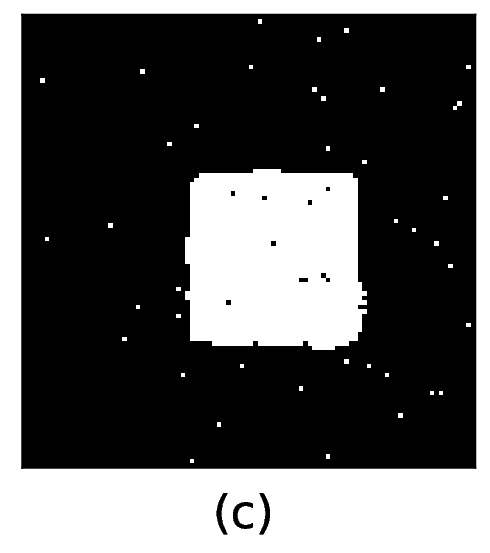}
		\includegraphics[width=0.35\linewidth]{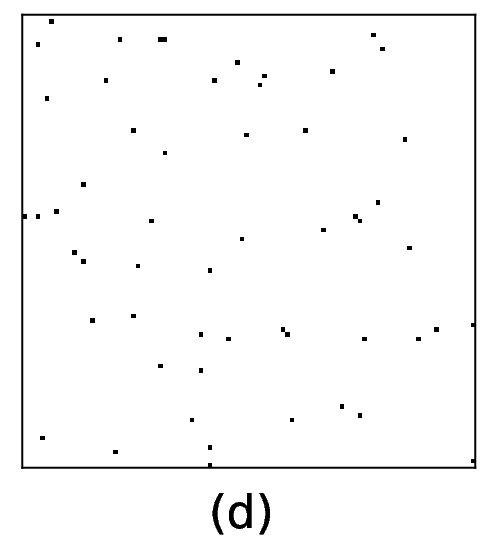}
		\caption{ 
			\textbf{Evolution with prepared initial state.}
			The spatio-temporal pattern evolution starting from a round shape of cooperators, with time $t=0$, $3\times10^6$, $10^7$, $4\times10^7$ Monte Carlo steps, respectively for (a) to (d). Other settings are exactly the same as Fig. \ref{fig:snapshot}(b,c). 
			Parameters: $c= 0.5$, $p= 1$, $\rho=0.2$ and $b = 0.1$.}
		\label{fig:snapshot_round}
\end{figure}

With these observations, the nucleation process can then be understood as follows. For those who are surrounded by many defectors, they tend to choose defection, but those who are located in a cooperative neighborhood opt to cooperate. This explains why patched C or D patterns are formed in Figs.~\ref{fig:snapshot}(b,c). At the interface of C/D patches, players still prefer cooperation mostly since cooperation is conditioned by at least two or more cooperators (including itself) are seen in their vicinity, which is easily satisfied. For more clear illustration, Fig.~\ref{fig:snapshot_round} shows a spatiotemporal pattern evolution with a prepared initial configuration, where the cooperators are located in a patched round. As seen, the patch remains compact and expands all the time. As time goes by, straight lines are formed at their interface, but any fluctuations along the straight boundary convert the defectors into cooperators as they get two or more cooperative neighbors to trigger cooperation.  Random initial conditions yield jagged boundaries as in Figs.~\ref{fig:snapshot}(b,c), where the expansion of cooperator clusters is even faster. 
As a result, there is a proliferation of cooperation and dominates the population given compact cooperation clusters are formed. 
Otherwise, there is no possibility for cooperators to survive when all are within the state $s_0$ or $s_1$. 

In fact, the Q-table evolution for cases outside of the bistable region (e.g. Fig.~\ref{fig:snapshot}(a-d)) shows quite similar features as observed in Fig.~\ref{fig:Qtable}, with only an exception that the density of $s_0$ does not flourish.

\begin{figure}[tbp]
		\centering
		\includegraphics[width=0.9\linewidth]{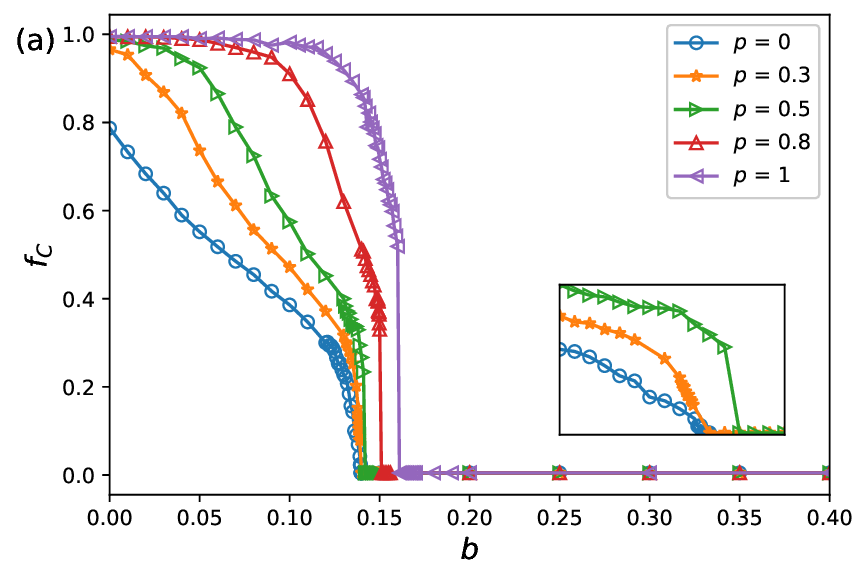} 
		\includegraphics[width=0.9\linewidth]{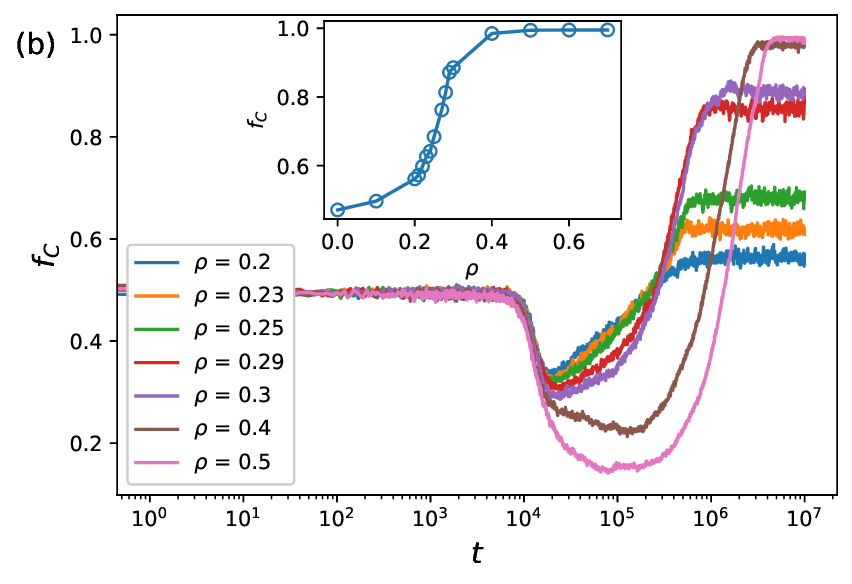}
		\caption{
		\textbf{The impact of two game parameters on cooperation.}
			 (a) The fraction of cooperators $f_C$ as a function of the temptation $b$ for different punishment intensities $p$ by fixing $c=0.2$. The inset is a blown-up for three transitions. (b) Time series of $f_c$ for a couple of punishment probabilities $\rho$, and the inset show the dependence of $f_c$ on $\rho$, where $c=0.3$ and $p=0.6$.
			  Other parameters: $b=0.1$.
		}
		\label{fig:b}
\end{figure}


\section{5. The impact of game parameters}\label{sec:gamepara}

Finally, we investigate the impact of two game parameters on the cooperation level -- the temptation $b$ and punishment probability $\rho$, shown in Fig.~\ref{fig:b}. Fig.~\ref{fig:b}(a) gives the phase transitions of $f_C$ versus $b$ for a couple of costs $c$ by fixing the punishment intensity $p=1$.  As expected, the cooperation level $f_C$ decreases as the temptation $b$ is increased. The shown phase transitions are all of discontinuous type and a higher cost leads to a smaller threshold for being almost full defection, and a large gap in the discontinuity.  

Similarly, Fig.~\ref{fig:b}(b) shows the impact of punishment probability $\rho$ by fixing $c=0.2$ and $p=0.6$. It shows that a larger value of $\rho$ generally leads to a higher level of cooperation. This trend is in line with our expectation since a larger punishment probability makes the defectors more likely to be punished to force them turn to cooperation, playing the similar role of punishment intensity $p$.

\section{6. Conclusion} \label{sec:conclusion}
	
In this work, we have addressed an important question from a new perspective about the impact of punishment on the evolution of cooperation.
We have revealed an intriguing dependence of the cooperation level on the punishment intensity, the cost, and the game parameter etc, where both continuous and discontinuous phase transition are possible.
	
In fact, some previous works also reveal both types of cooperation transition when punishment is posed~\cite{Helbing2010Punish,Helbing2010Evolutionary, Szolnoki2013Correlation, Lucas2021Symbiotic}. For example, Ref.~\cite{Helbing2010Punish} reported that both first-order and second order phase transitions are seen when the punishment intensity or cost is varied. The fundamental difference lies in that fact that they all are within the imitation framework where individuals revise their strategies according to the Fermi rule, and instead the RL framework is adopted in our work, leading to different physics picture behind. 

From the practice point of view, great care needs to be taken since the resulting cooperation prevalence could be lower than the benchmark level as shown in Fig.~\ref{fig:p}(a). If the penalty is weak, punishment is not suggested to conduct. This is especially true when the cost is too high, where all players turn to defect and the cooperation level is greatly reduced compared to the benchmark case where no punishment is posed. 
	
There is no doubt that the proposed reinforcement learning provides insights into the impact of punishment on cooperation. The strength of the RL paradigm is that individuals are allowed to adaptively adjust their moves to maximize their accumulated rewards according to their surrounding changes, which is what many creatures are doing in this world. At the game-theoretic side,  the RL has applied to understand some puzzles in socio-economic actives, such as the emergence of trust, the efficient resource allocation, etc. At the experimental side, we hope related behavioral experiments can be performed to justify or falsify our findings~\cite{Huang2015Experimental}, and more importantly the rationale behind the RL.

\section{Acknowledgments}
We acknowledge the X-Physics training program for undergraduate students in SNNU that initialized this research.
This work was supported by the National Natural Science Foundation of China [Grants Nos. 12075144,12165014]. ZGZ is supported by Excellent Graduate Training Program of Shaanxi Normal University [Grants No. LHRCTS23064]. ZC is supported by the Natural Science Basic Research Plan in Shaanxi Province of China (Grant No. 2023-JC-QN-0013) and Scientific Research Program Funded by Shaanxi Provincial Education Department (Program No. 23JK0604).
	
\bibliography{References}

\begin{thebibliography}{69}%
\makeatletter
\providecommand \@ifxundefined [1]{%
 \@ifx{#1\undefined}
}%
\providecommand \@ifnum [1]{%
 \ifnum #1\expandafter \@firstoftwo
 \else \expandafter \@secondoftwo
 \fi
}%
\providecommand \@ifx [1]{%
 \ifx #1\expandafter \@firstoftwo
 \else \expandafter \@secondoftwo
 \fi
}%
\providecommand \natexlab [1]{#1}%
\providecommand \enquote  [1]{``#1''}%
\providecommand \bibnamefont  [1]{#1}%
\providecommand \bibfnamefont [1]{#1}%
\providecommand \citenamefont [1]{#1}%
\providecommand \href@noop [0]{\@secondoftwo}%
\providecommand \href [0]{\begingroup \@sanitize@url \@href}%
\providecommand \@href[1]{\@@startlink{#1}\@@href}%
\providecommand \@@href[1]{\endgroup#1\@@endlink}%
\providecommand \@sanitize@url [0]{\catcode `\\12\catcode `\$12\catcode
  `\&12\catcode `\#12\catcode `\^12\catcode `\_12\catcode `\%12\relax}%
\providecommand \@@startlink[1]{}%
\providecommand \@@endlink[0]{}%
\providecommand \url  [0]{\begingroup\@sanitize@url \@url }%
\providecommand \@url [1]{\endgroup\@href {#1}{\urlprefix }}%
\providecommand \urlprefix  [0]{URL }%
\providecommand \Eprint [0]{\href }%
\providecommand \doibase [0]{http://dx.doi.org/}%
\providecommand \selectlanguage [0]{\@gobble}%
\providecommand \bibinfo  [0]{\@secondoftwo}%
\providecommand \bibfield  [0]{\@secondoftwo}%
\providecommand \translation [1]{[#1]}%
\providecommand \BibitemOpen [0]{}%
\providecommand \bibitemStop [0]{}%
\providecommand \bibitemNoStop [0]{.\EOS\space}%
\providecommand \EOS [0]{\spacefactor3000\relax}%
\providecommand \BibitemShut  [1]{\csname bibitem#1\endcsname}%
\let\auto@bib@innerbib\@empty
\bibitem [{\citenamefont {Nielsen}(1985)}]{Nielsen1985cooperation}%
  \BibitemOpen
  \bibfield  {author} {\bibinfo {author} {\bibfnamefont {R.~P.}\ \bibnamefont
  {Nielsen}},\ }\href {\doibase 10.2307/257983} {\bibfield  {journal} {\bibinfo
   {journal} {The Academy of Management Review}\ }\textbf {\bibinfo {volume}
  {10}},\ \bibinfo {pages} {368} (\bibinfo {year} {1985})}\BibitemShut
  {NoStop}%
\bibitem [{\citenamefont {Colman}(1995)}]{Colman1995Theory}%
  \BibitemOpen
  \bibfield  {author} {\bibinfo {author} {\bibfnamefont {A.}~\bibnamefont
  {Colman}},\ }\href {\doibase 10.4324/9780203761335} {\emph {\bibinfo {title}
  {Game Theory and Its Applications in the Social and Biological Sciences}}}\
  (\bibinfo  {publisher} {Psychology Press},\ \bibinfo {year}
  {1995})\BibitemShut {NoStop}%
\bibitem [{\citenamefont {Nowak}(2006{\natexlab{a}})}]{Nowak2006Evolution}%
  \BibitemOpen
  \bibfield  {author} {\bibinfo {author} {\bibfnamefont {M.~A.}\ \bibnamefont
  {Nowak}},\ }\href {https://www.hup.harvard.edu/books/9780674023383} {\emph
  {\bibinfo {title} {Evolutionary Dynamics}}}\ (\bibinfo  {publisher}
  {Belknap/Harvard},\ \bibinfo {year} {2006})\BibitemShut {NoStop}%
\bibitem [{\citenamefont {Pennisi}(2005)}]{Pennisi2005How}%
  \BibitemOpen
  \bibfield  {author} {\bibinfo {author} {\bibfnamefont {E.}~\bibnamefont
  {Pennisi}},\ }\href {\doibase 10.1126/science.309.5731.93} {\bibfield
  {journal} {\bibinfo  {journal} {Science}\ }\textbf {\bibinfo {volume}
  {309}},\ \bibinfo {pages} {93} (\bibinfo {year} {2005})}\BibitemShut
  {NoStop}%
\bibitem [{\citenamefont {Nowak}(2006{\natexlab{b}})}]{Nowak2006Five}%
  \BibitemOpen
  \bibfield  {author} {\bibinfo {author} {\bibfnamefont {M.~A.}\ \bibnamefont
  {Nowak}},\ }\href {\doibase 10.1126/science.1133755} {\bibfield  {journal}
  {\bibinfo  {journal} {Science}\ }\textbf {\bibinfo {volume} {314}},\ \bibinfo
  {pages} {1560} (\bibinfo {year} {2006}{\natexlab{b}})}\BibitemShut {NoStop}%
\bibitem [{\citenamefont {Dawkins}(2006)}]{Dawkins2006selfish}%
  \BibitemOpen
  \bibfield  {author} {\bibinfo {author} {\bibfnamefont {R.}~\bibnamefont
  {Dawkins}},\ }\href {\doibase 10.1007/978-3-319-16999-6_1876-1} {\emph
  {\bibinfo {title} {The Selfish Gene}}}\ (\bibinfo  {publisher} {Oxford
  University Press},\ \bibinfo {address} {New York, US},\ \bibinfo {year}
  {2006})\BibitemShut {NoStop}%
\bibitem [{\citenamefont {Wilson}()}]{Wilson1975Sociobiology}%
  \BibitemOpen
  \bibfield  {author} {\bibinfo {author} {\bibfnamefont {E.~O.}\ \bibnamefont
  {Wilson}},\ }\href@noop {} {\emph {\bibinfo {title} {Sociobiology: The new
  synthesis}}}\ (\bibinfo  {publisher} {Belknap Press of Harvard U Press},\
  \bibinfo {address} {Oxford, England})\BibitemShut {NoStop}%
\bibitem [{\citenamefont {Pacheco}\ \emph {et~al.}(2008)\citenamefont
  {Pacheco}, \citenamefont {Traulsen}, \citenamefont {Ohtsuki},\ and\
  \citenamefont {Nowak}}]{Nowak2008Repeated}%
  \BibitemOpen
  \bibfield  {author} {\bibinfo {author} {\bibfnamefont {J.~M.}\ \bibnamefont
  {Pacheco}}, \bibinfo {author} {\bibfnamefont {A.}~\bibnamefont {Traulsen}},
  \bibinfo {author} {\bibfnamefont {H.}~\bibnamefont {Ohtsuki}}, \ and\
  \bibinfo {author} {\bibfnamefont {M.~A.}\ \bibnamefont {Nowak}},\ }\href
  {\doibase https://doi.org/10.1016/j.jtbi.2007.10.040} {\bibfield  {journal}
  {\bibinfo  {journal} {Journal of Theoretical Biology}\ }\textbf {\bibinfo
  {volume} {250}},\ \bibinfo {pages} {723} (\bibinfo {year}
  {2008})}\BibitemShut {NoStop}%
\bibitem [{\citenamefont {Wang}\ \emph
  {et~al.}(2013{\natexlab{a}})\citenamefont {Wang}, \citenamefont {Szolnoki},\
  and\ \citenamefont {Perc}}]{Perc2013Interdependent}%
  \BibitemOpen
  \bibfield  {author} {\bibinfo {author} {\bibfnamefont {Z.}~\bibnamefont
  {Wang}}, \bibinfo {author} {\bibfnamefont {A.}~\bibnamefont {Szolnoki}}, \
  and\ \bibinfo {author} {\bibfnamefont {M.}~\bibnamefont {Perc}},\ }\href
  {\doibase 10.1038/srep01183} {\bibfield  {journal} {\bibinfo  {journal}
  {Scientific Reports}\ }\textbf {\bibinfo {volume} {3}},\ \bibinfo {pages}
  {1183} (\bibinfo {year} {2013}{\natexlab{a}})}\BibitemShut {NoStop}%
\bibitem [{\citenamefont {Nowak}\ and\ \citenamefont
  {Sigmund}(1998)}]{Nowak1998indirect}%
  \BibitemOpen
  \bibfield  {author} {\bibinfo {author} {\bibfnamefont {M.~A.}\ \bibnamefont
  {Nowak}}\ and\ \bibinfo {author} {\bibfnamefont {K.}~\bibnamefont
  {Sigmund}},\ }\href {\doibase 10.1038/31225} {\bibfield  {journal} {\bibinfo
  {journal} {Nature}\ }\textbf {\bibinfo {volume} {393}},\ \bibinfo {pages}
  {573} (\bibinfo {year} {1998})}\BibitemShut {NoStop}%
\bibitem [{\citenamefont {Ohtsuki}\ and\ \citenamefont
  {Iwasa}(2006)}]{Ohtsuki2006indirect}%
  \BibitemOpen
  \bibfield  {author} {\bibinfo {author} {\bibfnamefont {H.}~\bibnamefont
  {Ohtsuki}}\ and\ \bibinfo {author} {\bibfnamefont {Y.}~\bibnamefont
  {Iwasa}},\ }\href {\doibase 10.1016/j.jtbi.2005.08.008} {\bibfield  {journal}
  {\bibinfo  {journal} {Journal of Theoretical Biology}\ }\textbf {\bibinfo
  {volume} {239}},\ \bibinfo {pages} {435} (\bibinfo {year}
  {2006})}\BibitemShut {NoStop}%
\bibitem [{\citenamefont {Nowak}\ and\ \citenamefont
  {May}(1992)}]{Nowak1992spatial}%
  \BibitemOpen
  \bibfield  {author} {\bibinfo {author} {\bibfnamefont {M.~A.}\ \bibnamefont
  {Nowak}}\ and\ \bibinfo {author} {\bibfnamefont {R.~M.}\ \bibnamefont
  {May}},\ }\href {\doibase 10.1038/359826a0} {\bibfield  {journal} {\bibinfo
  {journal} {Nature}\ }\textbf {\bibinfo {volume} {359}},\ \bibinfo {pages}
  {826} (\bibinfo {year} {1992})}\BibitemShut {NoStop}%
\bibitem [{\citenamefont {Szab{\'o}}\ and\ \citenamefont
  {T{\"{o}}ke}(1998)}]{Szabo1998Evolutionary}%
  \BibitemOpen
  \bibfield  {author} {\bibinfo {author} {\bibfnamefont {G.}~\bibnamefont
  {Szab{\'o}}}\ and\ \bibinfo {author} {\bibfnamefont {C.}~\bibnamefont
  {T{\"{o}}ke}},\ }\href {\doibase 10.1103/PhysRevE.58.69} {\bibfield
  {journal} {\bibinfo  {journal} {Physical Review E}\ }\textbf {\bibinfo
  {volume} {58}},\ \bibinfo {pages} {69} (\bibinfo {year} {1998})}\BibitemShut
  {NoStop}%
\bibitem [{\citenamefont {Wang}\ \emph
  {et~al.}(2013{\natexlab{b}})\citenamefont {Wang}, \citenamefont {Szolnoki},\
  and\ \citenamefont {Perc}}]{Wang2013Interdependent}%
  \BibitemOpen
  \bibfield  {author} {\bibinfo {author} {\bibfnamefont {Z.}~\bibnamefont
  {Wang}}, \bibinfo {author} {\bibfnamefont {A.}~\bibnamefont {Szolnoki}}, \
  and\ \bibinfo {author} {\bibfnamefont {M.}~\bibnamefont {Perc}},\ }\href
  {\doibase 10.1038/srep01183} {\bibfield  {journal} {\bibinfo  {journal}
  {Scientific Reports}\ }\textbf {\bibinfo {volume} {3}},\ \bibinfo {pages}
  {1183} (\bibinfo {year} {2013}{\natexlab{b}})}\BibitemShut {NoStop}%
\bibitem [{\citenamefont {Smith}(1964)}]{Smith1964Group}%
  \BibitemOpen
  \bibfield  {author} {\bibinfo {author} {\bibfnamefont {J.~M.}\ \bibnamefont
  {Smith}},\ }\href {\doibase 10.1038/2011145a0} {\bibfield  {journal}
  {\bibinfo  {journal} {Nature}\ }\textbf {\bibinfo {volume} {201}},\ \bibinfo
  {pages} {1145} (\bibinfo {year} {1964})}\BibitemShut {NoStop}%
\bibitem [{\citenamefont {Charlesworth}(2000)}]{Charlesworth2000Levels}%
  \BibitemOpen
  \bibfield  {author} {\bibinfo {author} {\bibfnamefont {B.}~\bibnamefont
  {Charlesworth}},\ }\href {\doibase 10.1046/j.1365-2540.2000.0726a.x}
  {\bibfield  {journal} {\bibinfo  {journal} {Heredity}\ }\textbf {\bibinfo
  {volume} {84}},\ \bibinfo {pages} {493} (\bibinfo {year} {2000})}\BibitemShut
  {NoStop}%
\bibitem [{\citenamefont {Antonioni}\ and\ \citenamefont
  {Cardillo}(2017)}]{Antonioni2017Coevolution}%
  \BibitemOpen
  \bibfield  {author} {\bibinfo {author} {\bibfnamefont {A.}~\bibnamefont
  {Antonioni}}\ and\ \bibinfo {author} {\bibfnamefont {A.}~\bibnamefont
  {Cardillo}},\ }\href {\doibase 10.1103/PhysRevLett.118.238301} {\bibfield
  {journal} {\bibinfo  {journal} {Physical Review Letters}\ }\textbf {\bibinfo
  {volume} {118}},\ \bibinfo {pages} {238301} (\bibinfo {year}
  {2017})}\BibitemShut {NoStop}%
\bibitem [{\citenamefont {Nowak}\ \emph {et~al.}(2004)\citenamefont {Nowak},
  \citenamefont {Sasaki}, \citenamefont {Taylor},\ and\ \citenamefont
  {Fudenberg}}]{Nowak2004Emergence}%
  \BibitemOpen
  \bibfield  {author} {\bibinfo {author} {\bibfnamefont {M.~A.}\ \bibnamefont
  {Nowak}}, \bibinfo {author} {\bibfnamefont {A.}~\bibnamefont {Sasaki}},
  \bibinfo {author} {\bibfnamefont {C.}~\bibnamefont {Taylor}}, \ and\ \bibinfo
  {author} {\bibfnamefont {D.}~\bibnamefont {Fudenberg}},\ }\href {\doibase
  10.1038/nature02414} {\bibfield  {journal} {\bibinfo  {journal} {Nature}\
  }\textbf {\bibinfo {volume} {428}},\ \bibinfo {pages} {646} (\bibinfo {year}
  {2004})}\BibitemShut {NoStop}%
\bibitem [{\citenamefont {Wang}\ \emph {et~al.}(2006)\citenamefont {Wang},
  \citenamefont {Ren}, \citenamefont {Chen},\ and\ \citenamefont
  {Wang}}]{Wang2006Memory}%
  \BibitemOpen
  \bibfield  {author} {\bibinfo {author} {\bibfnamefont {W.-X.}\ \bibnamefont
  {Wang}}, \bibinfo {author} {\bibfnamefont {J.}~\bibnamefont {Ren}}, \bibinfo
  {author} {\bibfnamefont {G.}~\bibnamefont {Chen}}, \ and\ \bibinfo {author}
  {\bibfnamefont {B.-H.}\ \bibnamefont {Wang}},\ }\href {\doibase
  10.1103/PhysRevE.74.056113} {\bibfield  {journal} {\bibinfo  {journal}
  {Physical Review E}\ }\textbf {\bibinfo {volume} {74}},\ \bibinfo {pages}
  {056113} (\bibinfo {year} {2006})}\BibitemShut {NoStop}%
\bibitem [{\citenamefont {Qin}\ \emph {et~al.}(2008)\citenamefont {Qin},
  \citenamefont {Chen}, \citenamefont {Zhao},\ and\ \citenamefont
  {Shi}}]{Qin2008Effect}%
  \BibitemOpen
  \bibfield  {author} {\bibinfo {author} {\bibfnamefont {S.-M.}\ \bibnamefont
  {Qin}}, \bibinfo {author} {\bibfnamefont {Y.}~\bibnamefont {Chen}}, \bibinfo
  {author} {\bibfnamefont {X.-Y.}\ \bibnamefont {Zhao}}, \ and\ \bibinfo
  {author} {\bibfnamefont {J.}~\bibnamefont {Shi}},\ }\href {\doibase
  10.1103/PhysRevE.78.041129} {\bibfield  {journal} {\bibinfo  {journal}
  {Physical Review E}\ }\textbf {\bibinfo {volume} {78}},\ \bibinfo {pages}
  {041129} (\bibinfo {year} {2008})}\BibitemShut {NoStop}%
\bibitem [{\citenamefont {Liu}\ \emph {et~al.}(2010)\citenamefont {Liu},
  \citenamefont {Li}, \citenamefont {Chen},\ and\ \citenamefont
  {Wang}}]{Liu2010Memory}%
  \BibitemOpen
  \bibfield  {author} {\bibinfo {author} {\bibfnamefont {Y.}~\bibnamefont
  {Liu}}, \bibinfo {author} {\bibfnamefont {Z.}~\bibnamefont {Li}}, \bibinfo
  {author} {\bibfnamefont {X.}~\bibnamefont {Chen}}, \ and\ \bibinfo {author}
  {\bibfnamefont {L.}~\bibnamefont {Wang}},\ }\href {\doibase
  https://doi.org/10.1016/j.physa.2010.02.008} {\bibfield  {journal} {\bibinfo
  {journal} {Physica A: Statistical Mechanics and its Applications}\ }\textbf
  {\bibinfo {volume} {389}},\ \bibinfo {pages} {2390} (\bibinfo {year}
  {2010})}\BibitemShut {NoStop}%
\bibitem [{\citenamefont {Zhang}\ \emph {et~al.}(2023)\citenamefont {Zhang},
  \citenamefont {Li}, \citenamefont {Zhang}, \citenamefont {Ma}, \citenamefont
  {Zheng},\ and\ \citenamefont {Chen}}]{Zhang2023emergence}%
  \BibitemOpen
  \bibfield  {author} {\bibinfo {author} {\bibfnamefont {J.}~\bibnamefont
  {Zhang}}, \bibinfo {author} {\bibfnamefont {Z.}~\bibnamefont {Li}}, \bibinfo
  {author} {\bibfnamefont {J.}~\bibnamefont {Zhang}}, \bibinfo {author}
  {\bibfnamefont {L.}~\bibnamefont {Ma}}, \bibinfo {author} {\bibfnamefont
  {G.}~\bibnamefont {Zheng}}, \ and\ \bibinfo {author} {\bibfnamefont
  {L.}~\bibnamefont {Chen}},\ }\href {\doibase
  https://doi.org/10.1016/j.physa.2023.128682} {\bibfield  {journal} {\bibinfo
  {journal} {Physica A}\ }\textbf {\bibinfo {volume} {617}},\ \bibinfo {pages}
  {128682} (\bibinfo {year} {2023})}\BibitemShut {NoStop}%
\bibitem [{\citenamefont {Perc}\ and\ \citenamefont
  {Szolnoki}(2008)}]{Perc2008Social}%
  \BibitemOpen
  \bibfield  {author} {\bibinfo {author} {\bibfnamefont {M.}~\bibnamefont
  {Perc}}\ and\ \bibinfo {author} {\bibfnamefont {A.}~\bibnamefont
  {Szolnoki}},\ }\href {\doibase 10.1103/PhysRevE.77.011904} {\bibfield
  {journal} {\bibinfo  {journal} {Physical Review E}\ }\textbf {\bibinfo
  {volume} {77}},\ \bibinfo {pages} {011904} (\bibinfo {year}
  {2008})}\BibitemShut {NoStop}%
\bibitem [{\citenamefont {Santos}\ \emph {et~al.}(2008)\citenamefont {Santos},
  \citenamefont {Santos},\ and\ \citenamefont {Pacheco}}]{Santos2008Social}%
  \BibitemOpen
  \bibfield  {author} {\bibinfo {author} {\bibfnamefont {F.~C.}\ \bibnamefont
  {Santos}}, \bibinfo {author} {\bibfnamefont {M.~D.}\ \bibnamefont {Santos}},
  \ and\ \bibinfo {author} {\bibfnamefont {J.~M.}\ \bibnamefont {Pacheco}},\
  }\href {\doibase 10.1038/nature06940} {\bibfield  {journal} {\bibinfo
  {journal} {Nature}\ }\textbf {\bibinfo {volume} {454}},\ \bibinfo {pages}
  {213} (\bibinfo {year} {2008})}\BibitemShut {NoStop}%
\bibitem [{\citenamefont {Liang}\ \emph {et~al.}(2021)\citenamefont {Liang},
  \citenamefont {Zhang}, \citenamefont {Zheng},\ and\ \citenamefont
  {Chen}}]{Liang2021Social}%
  \BibitemOpen
  \bibfield  {author} {\bibinfo {author} {\bibfnamefont {R.}~\bibnamefont
  {Liang}}, \bibinfo {author} {\bibfnamefont {J.}~\bibnamefont {Zhang}},
  \bibinfo {author} {\bibfnamefont {G.}~\bibnamefont {Zheng}}, \ and\ \bibinfo
  {author} {\bibfnamefont {L.}~\bibnamefont {Chen}},\ }\href {\doibase
  https://doi.org/10.1016/j.physa.2020.125726} {\bibfield  {journal} {\bibinfo
  {journal} {Physica A}\ }\textbf {\bibinfo {volume} {567}},\ \bibinfo {pages}
  {125726} (\bibinfo {year} {2021})}\BibitemShut {NoStop}%
\bibitem [{\citenamefont {Liang}\ \emph {et~al.}(2022)\citenamefont {Liang},
  \citenamefont {Wang}, \citenamefont {Zhang}, \citenamefont {Zheng},
  \citenamefont {Ma},\ and\ \citenamefont {Chen}}]{Liang2022dynamical}%
  \BibitemOpen
  \bibfield  {author} {\bibinfo {author} {\bibfnamefont {R.}~\bibnamefont
  {Liang}}, \bibinfo {author} {\bibfnamefont {Q.}~\bibnamefont {Wang}},
  \bibinfo {author} {\bibfnamefont {J.}~\bibnamefont {Zhang}}, \bibinfo
  {author} {\bibfnamefont {G.}~\bibnamefont {Zheng}}, \bibinfo {author}
  {\bibfnamefont {L.}~\bibnamefont {Ma}}, \ and\ \bibinfo {author}
  {\bibfnamefont {L.}~\bibnamefont {Chen}},\ }\href {\doibase
  https://doi.org/10.1103/PhysRevE.105.054302} {\bibfield  {journal} {\bibinfo
  {journal} {Physical Review E}\ }\textbf {\bibinfo {volume} {105}},\ \bibinfo
  {pages} {054302} (\bibinfo {year} {2022})}\BibitemShut {NoStop}%
\bibitem [{\citenamefont {Sigmund}\ \emph {et~al.}(2001)\citenamefont
  {Sigmund}, \citenamefont {Hauert},\ and\ \citenamefont
  {Nowak}}]{Sigmund2001Reward}%
  \BibitemOpen
  \bibfield  {author} {\bibinfo {author} {\bibfnamefont {K.}~\bibnamefont
  {Sigmund}}, \bibinfo {author} {\bibfnamefont {C.}~\bibnamefont {Hauert}}, \
  and\ \bibinfo {author} {\bibfnamefont {M.~A.}\ \bibnamefont {Nowak}},\ }\href
  {\doibase https://doi.org/10.1073/pnas.161155698} {\bibfield  {journal}
  {\bibinfo  {journal} {Proceedings of the National Academy of Sciences}\ }
  (\bibinfo {year} {2001}),\
  https://doi.org/10.1073/pnas.161155698}\BibitemShut {NoStop}%
\bibitem [{\citenamefont {Ma}\ \emph {et~al.}(2023)\citenamefont {Ma},
  \citenamefont {Zhang}, \citenamefont {Zheng}, \citenamefont {Liang},\ and\
  \citenamefont {Chen}}]{Ma2023emergence}%
  \BibitemOpen
  \bibfield  {author} {\bibinfo {author} {\bibfnamefont {L.}~\bibnamefont
  {Ma}}, \bibinfo {author} {\bibfnamefont {J.}~\bibnamefont {Zhang}}, \bibinfo
  {author} {\bibfnamefont {G.}~\bibnamefont {Zheng}}, \bibinfo {author}
  {\bibfnamefont {R.}~\bibnamefont {Liang}}, \ and\ \bibinfo {author}
  {\bibfnamefont {L.}~\bibnamefont {Chen}},\ }\href {\doibase
  https://doi.org/10.1016/j.chaos.2023.113452} {\bibfield  {journal} {\bibinfo
  {journal} {Chaos, Solitons \& Fractals}\ }\textbf {\bibinfo {volume} {171}},\
  \bibinfo {pages} {113452} (\bibinfo {year} {2023})}\BibitemShut {NoStop}%
\bibitem [{\citenamefont {Takesue}(2018)}]{Takesue2018Evolutionary}%
  \BibitemOpen
  \bibfield  {author} {\bibinfo {author} {\bibfnamefont {H.}~\bibnamefont
  {Takesue}},\ }\href {\doibase 10.1209/0295-5075/121/48005} {\bibfield
  {journal} {\bibinfo  {journal} {Europhysics Letters}\ }\textbf {\bibinfo
  {volume} {121}} (\bibinfo {year} {2018}),\
  10.1209/0295-5075/121/48005}\BibitemShut {NoStop}%
\bibitem [{\citenamefont {Szolnoki}\ and\ \citenamefont
  {Perc}(2013{\natexlab{a}})}]{Szolnoki2013Effectiveness}%
  \BibitemOpen
  \bibfield  {author} {\bibinfo {author} {\bibfnamefont {A.}~\bibnamefont
  {Szolnoki}}\ and\ \bibinfo {author} {\bibfnamefont {M.}~\bibnamefont
  {Perc}},\ }\href {\doibase 10.1016/j.jtbi.2013.02.008} {\bibfield  {journal}
  {\bibinfo  {journal} {Journal of Theoretical Biology}\ }\textbf {\bibinfo
  {volume} {325}},\ \bibinfo {pages} {34} (\bibinfo {year}
  {2013}{\natexlab{a}})}\BibitemShut {NoStop}%
\bibitem [{\citenamefont {Flores}\ \emph {et~al.}(2021)\citenamefont {Flores},
  \citenamefont {Fernandes}, \citenamefont {Amaral},\ and\ \citenamefont
  {Vainstein}}]{Lucas2021Symbiotic}%
  \BibitemOpen
  \bibfield  {author} {\bibinfo {author} {\bibfnamefont {L.~S.}\ \bibnamefont
  {Flores}}, \bibinfo {author} {\bibfnamefont {H.~C.}\ \bibnamefont
  {Fernandes}}, \bibinfo {author} {\bibfnamefont {M.~A.}\ \bibnamefont
  {Amaral}}, \ and\ \bibinfo {author} {\bibfnamefont {M.~H.}\ \bibnamefont
  {Vainstein}},\ }\href {\doibase https://doi.org/10.1016/j.jtbi.2021.110737}
  {\bibfield  {journal} {\bibinfo  {journal} {Journal of Theoretical Biology}\
  }\textbf {\bibinfo {volume} {524}},\ \bibinfo {pages} {110737} (\bibinfo
  {year} {2021})}\BibitemShut {NoStop}%
\bibitem [{\citenamefont {Herrmann}\ \emph {et~al.}(2008)\citenamefont
  {Herrmann}, \citenamefont {Th\"{o}ni},\ and\ \citenamefont
  {G{\"a}chter}}]{Herrmann2008Antisocial}%
  \BibitemOpen
  \bibfield  {author} {\bibinfo {author} {\bibfnamefont {B.}~\bibnamefont
  {Herrmann}}, \bibinfo {author} {\bibfnamefont {C.}~\bibnamefont {Th\"{o}ni}},
  \ and\ \bibinfo {author} {\bibfnamefont {S.}~\bibnamefont {G{\"a}chter}},\
  }\href {\doibase 10.1126/science.1153808} {\bibfield  {journal} {\bibinfo
  {journal} {Science}\ }\textbf {\bibinfo {volume} {319}},\ \bibinfo {pages}
  {1362} (\bibinfo {year} {2008})}\BibitemShut {NoStop}%
\bibitem [{\citenamefont {Diekmann}\ and\ \citenamefont
  {Przepiorka}(2015)}]{Diekmann2015Punitive}%
  \BibitemOpen
  \bibfield  {author} {\bibinfo {author} {\bibfnamefont {A.}~\bibnamefont
  {Diekmann}}\ and\ \bibinfo {author} {\bibfnamefont {W.}~\bibnamefont
  {Przepiorka}},\ }\href {\doibase 10.1038/srep10321} {\bibfield  {journal}
  {\bibinfo  {journal} {Scientific Reports}\ }\textbf {\bibinfo {volume} {5}},\
  \bibinfo {pages} {10321} (\bibinfo {year} {2015})}\BibitemShut {NoStop}%
\bibitem [{\citenamefont {Fehr}\ and\ \citenamefont
  {G\"{a}chter}(2002)}]{Fehr2002Altruistic}%
  \BibitemOpen
  \bibfield  {author} {\bibinfo {author} {\bibfnamefont {E.}~\bibnamefont
  {Fehr}}\ and\ \bibinfo {author} {\bibfnamefont {S.}~\bibnamefont
  {G\"{a}chter}},\ }\href {\doibase https://doi.org/10.1038/415137a} {\bibfield
   {journal} {\bibinfo  {journal} {Nature}\ }\textbf {\bibinfo {volume}
  {415}},\ \bibinfo {pages} {137–140} (\bibinfo {year} {2002})}\BibitemShut
  {NoStop}%
\bibitem [{\citenamefont {Henrich}\ \emph {et~al.}(2006)\citenamefont
  {Henrich}, \citenamefont {McElreath}, \citenamefont {Barr}, \citenamefont
  {Ensminger}, \citenamefont {Barrett}, \citenamefont {Bolyanatz},
  \citenamefont {Cardenas}, \citenamefont {Gurven}, \citenamefont {Gwako},
  \citenamefont {Henrich}, \citenamefont {Lesorogol}, \citenamefont {Marlowe},
  \citenamefont {Tracer},\ and\ \citenamefont {Ziker}}]{Henrich2006Costly}%
  \BibitemOpen
  \bibfield  {author} {\bibinfo {author} {\bibfnamefont {J.}~\bibnamefont
  {Henrich}}, \bibinfo {author} {\bibfnamefont {R.}~\bibnamefont {McElreath}},
  \bibinfo {author} {\bibfnamefont {A.}~\bibnamefont {Barr}}, \bibinfo {author}
  {\bibfnamefont {J.}~\bibnamefont {Ensminger}}, \bibinfo {author}
  {\bibfnamefont {C.}~\bibnamefont {Barrett}}, \bibinfo {author} {\bibfnamefont
  {A.}~\bibnamefont {Bolyanatz}}, \bibinfo {author} {\bibfnamefont {J.~C.}\
  \bibnamefont {Cardenas}}, \bibinfo {author} {\bibfnamefont {M.}~\bibnamefont
  {Gurven}}, \bibinfo {author} {\bibfnamefont {E.}~\bibnamefont {Gwako}},
  \bibinfo {author} {\bibfnamefont {N.}~\bibnamefont {Henrich}}, \bibinfo
  {author} {\bibfnamefont {C.}~\bibnamefont {Lesorogol}}, \bibinfo {author}
  {\bibfnamefont {F.}~\bibnamefont {Marlowe}}, \bibinfo {author} {\bibfnamefont
  {D.}~\bibnamefont {Tracer}}, \ and\ \bibinfo {author} {\bibfnamefont
  {J.}~\bibnamefont {Ziker}},\ }\href {\doibase doi:10.1126/science.1127333}
  {\bibfield  {journal} {\bibinfo  {journal} {Science}\ }\textbf {\bibinfo
  {volume} {312}},\ \bibinfo {pages} {1767} (\bibinfo {year}
  {2006})}\BibitemShut {NoStop}%
\bibitem [{\citenamefont {Sigmund}(2007)}]{Sigmund2007Punish}%
  \BibitemOpen
  \bibfield  {author} {\bibinfo {author} {\bibfnamefont {K.}~\bibnamefont
  {Sigmund}},\ }\href {\doibase https://doi.org/10.1016/j.tree.2007.06.012}
  {\bibfield  {journal} {\bibinfo  {journal} {Trends in Ecology \& Evolution}\
  }\textbf {\bibinfo {volume} {22}},\ \bibinfo {pages} {593} (\bibinfo {year}
  {2007})}\BibitemShut {NoStop}%
\bibitem [{\citenamefont {Brandt}\ \emph {et~al.}(2003)\citenamefont {Brandt},
  \citenamefont {Hauert},\ and\ \citenamefont
  {Sigmund}}]{Brandt2003Punishment}%
  \BibitemOpen
  \bibfield  {author} {\bibinfo {author} {\bibfnamefont {H.}~\bibnamefont
  {Brandt}}, \bibinfo {author} {\bibfnamefont {C.}~\bibnamefont {Hauert}}, \
  and\ \bibinfo {author} {\bibfnamefont {K.}~\bibnamefont {Sigmund}},\ }\href
  {\doibase 10.1098/rspb.2003.2336} {\bibfield  {journal} {\bibinfo  {journal}
  {Proceedings of the Royal Society B}\ }\textbf {\bibinfo {volume} {270}},\
  \bibinfo {pages} {1099} (\bibinfo {year} {2003})}\BibitemShut {NoStop}%
\bibitem [{\citenamefont {Yang}\ \emph {et~al.}(2015)\citenamefont {Yang},
  \citenamefont {Wu}, \citenamefont {Rong},\ and\ \citenamefont
  {Lai}}]{Yang2015Peer}%
  \BibitemOpen
  \bibfield  {author} {\bibinfo {author} {\bibfnamefont {H.~X.}\ \bibnamefont
  {Yang}}, \bibinfo {author} {\bibfnamefont {Z.~X.}\ \bibnamefont {Wu}},
  \bibinfo {author} {\bibfnamefont {Z.}~\bibnamefont {Rong}}, \ and\ \bibinfo
  {author} {\bibfnamefont {Y.~C.}\ \bibnamefont {Lai}},\ }\href {\doibase
  10.1103/PhysRevE.91.022121} {\bibfield  {journal} {\bibinfo  {journal}
  {Physical Review E}\ }\textbf {\bibinfo {volume} {91}},\ \bibinfo {pages}
  {022121} (\bibinfo {year} {2015})}\BibitemShut {NoStop}%
\bibitem [{\citenamefont {Sigmund}\ \emph {et~al.}(2010)\citenamefont
  {Sigmund}, \citenamefont {De~Silva}, \citenamefont {Traulsen},\ and\
  \citenamefont {Hauert}}]{Sigmund2010Social}%
  \BibitemOpen
  \bibfield  {author} {\bibinfo {author} {\bibfnamefont {K.}~\bibnamefont
  {Sigmund}}, \bibinfo {author} {\bibfnamefont {H.}~\bibnamefont {De~Silva}},
  \bibinfo {author} {\bibfnamefont {A.}~\bibnamefont {Traulsen}}, \ and\
  \bibinfo {author} {\bibfnamefont {C.}~\bibnamefont {Hauert}},\ }\href
  {\doibase 10.1038/nature09203} {\bibfield  {journal} {\bibinfo  {journal}
  {Nature}\ }\textbf {\bibinfo {volume} {466}},\ \bibinfo {pages} {861}
  (\bibinfo {year} {2010})}\BibitemShut {NoStop}%
\bibitem [{\citenamefont {Szolnoki}\ \emph {et~al.}(2011)\citenamefont
  {Szolnoki}, \citenamefont {Szab\'{o}},\ and\ \citenamefont
  {Perc}}]{Szolnoki2011Phase}%
  \BibitemOpen
  \bibfield  {author} {\bibinfo {author} {\bibfnamefont {A.}~\bibnamefont
  {Szolnoki}}, \bibinfo {author} {\bibfnamefont {G.}~\bibnamefont {Szab\'{o}}},
  \ and\ \bibinfo {author} {\bibfnamefont {M.}~\bibnamefont {Perc}},\ }\href
  {\doibase 10.1103/PhysRevE.83.036101} {\bibfield  {journal} {\bibinfo
  {journal} {Physical Review E}\ }\textbf {\bibinfo {volume} {83}},\ \bibinfo
  {pages} {036101} (\bibinfo {year} {2011})}\BibitemShut {NoStop}%
\bibitem [{\citenamefont {Perc}\ \emph {et~al.}(2017)\citenamefont {Perc},
  \citenamefont {Jordan}, \citenamefont {Rand}, \citenamefont {Wang},
  \citenamefont {Boccaletti},\ and\ \citenamefont
  {Szolnoki}}]{Perc2017Statistical}%
  \BibitemOpen
  \bibfield  {author} {\bibinfo {author} {\bibfnamefont {M.}~\bibnamefont
  {Perc}}, \bibinfo {author} {\bibfnamefont {J.~J.}\ \bibnamefont {Jordan}},
  \bibinfo {author} {\bibfnamefont {D.~G.}\ \bibnamefont {Rand}}, \bibinfo
  {author} {\bibfnamefont {Z.}~\bibnamefont {Wang}}, \bibinfo {author}
  {\bibfnamefont {S.}~\bibnamefont {Boccaletti}}, \ and\ \bibinfo {author}
  {\bibfnamefont {A.}~\bibnamefont {Szolnoki}},\ }\href {\doibase
  10.1016/j.physrep.2017.05.004} {\bibfield  {journal} {\bibinfo  {journal}
  {Physics Reports}\ }\textbf {\bibinfo {volume} {687}},\ \bibinfo {pages} {1}
  (\bibinfo {year} {2017})}\BibitemShut {NoStop}%
\bibitem [{\citenamefont {Amor}\ and\ \citenamefont
  {Fort}(2011)}]{Amor2011Effects}%
  \BibitemOpen
  \bibfield  {author} {\bibinfo {author} {\bibfnamefont {D.~R.}\ \bibnamefont
  {Amor}}\ and\ \bibinfo {author} {\bibfnamefont {J.}~\bibnamefont {Fort}},\
  }\href {\doibase 10.1103/PhysRevE.84.066115} {\bibfield  {journal} {\bibinfo
  {journal} {Physical Review E}\ }\textbf {\bibinfo {volume} {84}},\ \bibinfo
  {pages} {066115} (\bibinfo {year} {2011})}\BibitemShut {NoStop}%
\bibitem [{\citenamefont {Boyd}\ \emph {et~al.}(2003)\citenamefont {Boyd},
  \citenamefont {Gintis}, \citenamefont {Bowles},\ and\ \citenamefont
  {Richerson}}]{Boyd2003evolution}%
  \BibitemOpen
  \bibfield  {author} {\bibinfo {author} {\bibfnamefont {R.}~\bibnamefont
  {Boyd}}, \bibinfo {author} {\bibfnamefont {H.}~\bibnamefont {Gintis}},
  \bibinfo {author} {\bibfnamefont {S.}~\bibnamefont {Bowles}}, \ and\ \bibinfo
  {author} {\bibfnamefont {P.~J.}\ \bibnamefont {Richerson}},\ }\href {\doibase
  doi:10.1073/pnas.0630443100} {\bibfield  {journal} {\bibinfo  {journal}
  {Proceedings of the National Academy of Sciences}\ }\textbf {\bibinfo
  {volume} {100}},\ \bibinfo {pages} {3531} (\bibinfo {year}
  {2003})}\BibitemShut {NoStop}%
\bibitem [{\citenamefont {Yang}\ and\ \citenamefont
  {Wang}(2015)}]{Yang2015Role}%
  \BibitemOpen
  \bibfield  {author} {\bibinfo {author} {\bibfnamefont {H.-X.}\ \bibnamefont
  {Yang}}\ and\ \bibinfo {author} {\bibfnamefont {Z.}~\bibnamefont {Wang}},\
  }\href {\doibase 10.1209/0295-5075/111/60003} {\bibfield  {journal} {\bibinfo
   {journal} {Europhysics Letters}\ }\textbf {\bibinfo {volume} {111}},\
  \bibinfo {pages} {60003} (\bibinfo {year} {2015})}\BibitemShut {NoStop}%
\bibitem [{\citenamefont {Yang}\ \emph {et~al.}(2017)\citenamefont {Yang},
  \citenamefont {Fan}, \citenamefont {Liu},\ and\ \citenamefont
  {Chen}}]{Yang2017Phase}%
  \BibitemOpen
  \bibfield  {author} {\bibinfo {author} {\bibfnamefont {B.}~\bibnamefont
  {Yang}}, \bibinfo {author} {\bibfnamefont {M.}~\bibnamefont {Fan}}, \bibinfo
  {author} {\bibfnamefont {W.-Q.}\ \bibnamefont {Liu}}, \ and\ \bibinfo
  {author} {\bibfnamefont {X.-S.}\ \bibnamefont {Chen}},\ }\href {\doibase
  10.7498/aps.66.196401} {\bibfield  {journal} {\bibinfo  {journal} {Acta
  Physica Sinica}\ }\textbf {\bibinfo {volume} {66}},\ \bibinfo {pages}
  {196401} (\bibinfo {year} {2017})}\BibitemShut {NoStop}%
\bibitem [{\citenamefont {Helbing}\ \emph
  {et~al.}(2010{\natexlab{a}})\citenamefont {Helbing}, \citenamefont
  {Szolnoki}, \citenamefont {Perc},\ and\ \citenamefont
  {Szab\'{o}}}]{Helbing2010Punish}%
  \BibitemOpen
  \bibfield  {author} {\bibinfo {author} {\bibfnamefont {D.}~\bibnamefont
  {Helbing}}, \bibinfo {author} {\bibfnamefont {A.}~\bibnamefont {Szolnoki}},
  \bibinfo {author} {\bibfnamefont {M.}~\bibnamefont {Perc}}, \ and\ \bibinfo
  {author} {\bibfnamefont {G.}~\bibnamefont {Szab\'{o}}},\ }\href {\doibase
  10.1088/1367-2630/12/8/083005} {\bibfield  {journal} {\bibinfo  {journal}
  {New Journal of Physics}\ }\textbf {\bibinfo {volume} {12}},\ \bibinfo
  {pages} {083005} (\bibinfo {year} {2010}{\natexlab{a}})}\BibitemShut
  {NoStop}%
\bibitem [{\citenamefont {Helbing}\ \emph
  {et~al.}(2010{\natexlab{b}})\citenamefont {Helbing}, \citenamefont
  {Szolnoki}, \citenamefont {Perc},\ and\ \citenamefont
  {Szab\'{o}}}]{Helbing2010Evolutionary}%
  \BibitemOpen
  \bibfield  {author} {\bibinfo {author} {\bibfnamefont {D.}~\bibnamefont
  {Helbing}}, \bibinfo {author} {\bibfnamefont {A.}~\bibnamefont {Szolnoki}},
  \bibinfo {author} {\bibfnamefont {M.}~\bibnamefont {Perc}}, \ and\ \bibinfo
  {author} {\bibfnamefont {G.}~\bibnamefont {Szab\'{o}}},\ }\href {\doibase
  10.1371/journal.pcbi.1000758} {\bibfield  {journal} {\bibinfo  {journal}
  {PLoS Computional Biology}\ }\textbf {\bibinfo {volume} {6}},\ \bibinfo
  {pages} {e1000758} (\bibinfo {year} {2010}{\natexlab{b}})}\BibitemShut
  {NoStop}%
\bibitem [{\citenamefont {Helbing}\ \emph
  {et~al.}(2010{\natexlab{c}})\citenamefont {Helbing}, \citenamefont
  {Szolnoki}, \citenamefont {Perc},\ and\ \citenamefont
  {Szab\'{o}}}]{Helbing2010Defector}%
  \BibitemOpen
  \bibfield  {author} {\bibinfo {author} {\bibfnamefont {D.}~\bibnamefont
  {Helbing}}, \bibinfo {author} {\bibfnamefont {A.}~\bibnamefont {Szolnoki}},
  \bibinfo {author} {\bibfnamefont {M.}~\bibnamefont {Perc}}, \ and\ \bibinfo
  {author} {\bibfnamefont {G.}~\bibnamefont {Szab\'{o}}},\ }\href {\doibase
  10.1103/PhysRevE.81.057104} {\bibfield  {journal} {\bibinfo  {journal}
  {Physical Review E}\ }\textbf {\bibinfo {volume} {81}},\ \bibinfo {pages}
  {057104} (\bibinfo {year} {2010}{\natexlab{c}})}\BibitemShut {NoStop}%
\bibitem [{\citenamefont {Lv}\ and\ \citenamefont
  {Song}(2022)}]{LV2022Particle}%
  \BibitemOpen
  \bibfield  {author} {\bibinfo {author} {\bibfnamefont {S.}~\bibnamefont
  {Lv}}\ and\ \bibinfo {author} {\bibfnamefont {F.}~\bibnamefont {Song}},\
  }\href {\doibase https://doi.org/10.1016/j.amc.2021.126586} {\bibfield
  {journal} {\bibinfo  {journal} {Applied Mathematics and Computation}\
  }\textbf {\bibinfo {volume} {412}},\ \bibinfo {pages} {126586} (\bibinfo
  {year} {2022})}\BibitemShut {NoStop}%
\bibitem [{\citenamefont {Szolnoki}\ and\ \citenamefont
  {Perc}(2013{\natexlab{b}})}]{Szolnoki2013Correlation}%
  \BibitemOpen
  \bibfield  {author} {\bibinfo {author} {\bibfnamefont {A.}~\bibnamefont
  {Szolnoki}}\ and\ \bibinfo {author} {\bibfnamefont {M.}~\bibnamefont
  {Perc}},\ }\href {\doibase 10.1103/PhysRevX.3.041021} {\bibfield  {journal}
  {\bibinfo  {journal} {Physical Review X}\ }\textbf {\bibinfo {volume} {3}},\
  \bibinfo {pages} {041021} (\bibinfo {year} {2013}{\natexlab{b}})}\BibitemShut
  {NoStop}%
\bibitem [{\citenamefont {Sasaki}\ \emph {et~al.}(2015)\citenamefont {Sasaki},
  \citenamefont {Uchida},\ and\ \citenamefont {Chen}}]{Sasaki2015Voluntary}%
  \BibitemOpen
  \bibfield  {author} {\bibinfo {author} {\bibfnamefont {T.}~\bibnamefont
  {Sasaki}}, \bibinfo {author} {\bibfnamefont {S.}~\bibnamefont {Uchida}}, \
  and\ \bibinfo {author} {\bibfnamefont {X.}~\bibnamefont {Chen}},\ }\href
  {\doibase 10.1038/srep08917} {\bibfield  {journal} {\bibinfo  {journal}
  {Scientific Reports}\ }\textbf {\bibinfo {volume} {5}},\ \bibinfo {pages}
  {8917} (\bibinfo {year} {2015})}\BibitemShut {NoStop}%
\bibitem [{\citenamefont {Roca}\ \emph {et~al.}(2009)\citenamefont {Roca},
  \citenamefont {Cuesta},\ and\ \citenamefont
  {Sánchez}}]{Roca2009Evolutionary}%
  \BibitemOpen
  \bibfield  {author} {\bibinfo {author} {\bibfnamefont {C.~P.}\ \bibnamefont
  {Roca}}, \bibinfo {author} {\bibfnamefont {J.~A.}\ \bibnamefont {Cuesta}}, \
  and\ \bibinfo {author} {\bibfnamefont {A.}~\bibnamefont {Sánchez}},\ }\href
  {\doibase https://doi.org/10.1016/j.plrev.2009.08.001} {\bibfield  {journal}
  {\bibinfo  {journal} {Physics of Life Reviews}\ }\textbf {\bibinfo {volume}
  {6}},\ \bibinfo {pages} {208} (\bibinfo {year} {2009})}\BibitemShut {NoStop}%
\bibitem [{\citenamefont {Szolnoki}\ \emph {et~al.}(2009)\citenamefont
  {Szolnoki}, \citenamefont {Perc},\ and\ \citenamefont
  {Szabó}}]{Szolnoki2009Topology}%
  \BibitemOpen
  \bibfield  {author} {\bibinfo {author} {\bibfnamefont {A.}~\bibnamefont
  {Szolnoki}}, \bibinfo {author} {\bibfnamefont {M.}~\bibnamefont {Perc}}, \
  and\ \bibinfo {author} {\bibfnamefont {G.}~\bibnamefont {Szabó}},\ }\href
  {\doibase 10.1103/PhysRevE.80.056109} {\bibfield  {journal} {\bibinfo
  {journal} {Physical Review E}\ }\textbf {\bibinfo {volume} {80}},\ \bibinfo
  {pages} {056109} (\bibinfo {year} {2009})}\BibitemShut {NoStop}%
\bibitem [{\citenamefont {Bandura}\ and\ \citenamefont
  {Walters}(1977)}]{Bandura1977social}%
  \BibitemOpen
  \bibfield  {author} {\bibinfo {author} {\bibfnamefont {A.}~\bibnamefont
  {Bandura}}\ and\ \bibinfo {author} {\bibfnamefont {R.~H.}\ \bibnamefont
  {Walters}},\ }\href
  {https://link.springer.com/chapter/10.1007/978-3-030-43620-9_7} {\emph
  {\bibinfo {title} {Social Learning Theory}}},\ Vol.~\bibinfo {volume} {1}\
  (\bibinfo  {publisher} {Englewood cliffs Prentice Hall},\ \bibinfo {year}
  {1977})\BibitemShut {NoStop}%
\bibitem [{\citenamefont {Kaelbling}\ \emph {et~al.}(1996)\citenamefont
  {Kaelbling}, \citenamefont {Littman},\ and\ \citenamefont
  {Moore}}]{Kaelbling1996Reinforcement}%
  \BibitemOpen
  \bibfield  {author} {\bibinfo {author} {\bibfnamefont {L.~P.}\ \bibnamefont
  {Kaelbling}}, \bibinfo {author} {\bibfnamefont {M.~L.}\ \bibnamefont
  {Littman}}, \ and\ \bibinfo {author} {\bibfnamefont {A.~W.}\ \bibnamefont
  {Moore}},\ }\href {\doibase https://doi.org/10.1613/jair.301} {\bibfield
  {journal} {\bibinfo  {journal} {Journal of Artificial Intelligence Research}\
  }\textbf {\bibinfo {volume} {4}},\ \bibinfo {pages} {237} (\bibinfo {year}
  {1996})}\BibitemShut {NoStop}%
\bibitem [{\citenamefont {Watkins}\ and\ \citenamefont
  {Dayan}(1992)}]{Watkins1992Q-learning}%
  \BibitemOpen
  \bibfield  {author} {\bibinfo {author} {\bibfnamefont {C.~J. C.~H.}\
  \bibnamefont {Watkins}}\ and\ \bibinfo {author} {\bibfnamefont
  {P.}~\bibnamefont {Dayan}},\ }\href {\doibase 10.1007/BF00992698} {\bibfield
  {journal} {\bibinfo  {journal} {Machine Learning}\ }\textbf {\bibinfo
  {volume} {8}},\ \bibinfo {pages} {279} (\bibinfo {year} {1992})}\BibitemShut
  {NoStop}%
\bibitem [{\citenamefont {Zhang}\ \emph {et~al.}(2020)\citenamefont {Zhang},
  \citenamefont {Zhang}, \citenamefont {Chen},\ and\ \citenamefont
  {Liu}}]{Zhang2020Understanding}%
  \BibitemOpen
  \bibfield  {author} {\bibinfo {author} {\bibfnamefont {J.-Q.}\ \bibnamefont
  {Zhang}}, \bibinfo {author} {\bibfnamefont {S.-P.}\ \bibnamefont {Zhang}},
  \bibinfo {author} {\bibfnamefont {L.}~\bibnamefont {Chen}}, \ and\ \bibinfo
  {author} {\bibfnamefont {X.-D.}\ \bibnamefont {Liu}},\ }\href {\doibase
  10.1103/PhysRevE.101.042402} {\bibfield  {journal} {\bibinfo  {journal}
  {Physical Review E}\ }\textbf {\bibinfo {volume} {101}},\ \bibinfo {pages}
  {042402} (\bibinfo {year} {2020})}\BibitemShut {NoStop}%
\bibitem [{\citenamefont {Song}\ \emph {et~al.}(2022)\citenamefont {Song},
  \citenamefont {Guo}, \citenamefont {Jia}, \citenamefont {Perc}, \citenamefont
  {Li},\ and\ \citenamefont {Wang}}]{Zhao2022Reinforcement}%
  \BibitemOpen
  \bibfield  {author} {\bibinfo {author} {\bibfnamefont {Z.}~\bibnamefont
  {Song}}, \bibinfo {author} {\bibfnamefont {H.}~\bibnamefont {Guo}}, \bibinfo
  {author} {\bibfnamefont {D.}~\bibnamefont {Jia}}, \bibinfo {author}
  {\bibfnamefont {M.}~\bibnamefont {Perc}}, \bibinfo {author} {\bibfnamefont
  {X.}~\bibnamefont {Li}}, \ and\ \bibinfo {author} {\bibfnamefont
  {Z.}~\bibnamefont {Wang}},\ }\href {\doibase
  https://doi.org/10.1016/j.neucom.2022.09.109} {\bibfield  {journal} {\bibinfo
   {journal} {Neurocomputing}\ }\textbf {\bibinfo {volume} {513}},\ \bibinfo
  {pages} {104} (\bibinfo {year} {2022})}\BibitemShut {NoStop}%
\bibitem [{\citenamefont {Wang}\ \emph {et~al.}(2022)\citenamefont {Wang},
  \citenamefont {Jia}, \citenamefont {Zhang}, \citenamefont {Zhu},
  \citenamefont {Perc}, \citenamefont {Shi},\ and\ \citenamefont
  {Wang}}]{Wang2022Levy}%
  \BibitemOpen
  \bibfield  {author} {\bibinfo {author} {\bibfnamefont {L.}~\bibnamefont
  {Wang}}, \bibinfo {author} {\bibfnamefont {D.}~\bibnamefont {Jia}}, \bibinfo
  {author} {\bibfnamefont {L.}~\bibnamefont {Zhang}}, \bibinfo {author}
  {\bibfnamefont {P.}~\bibnamefont {Zhu}}, \bibinfo {author} {\bibfnamefont
  {M.}~\bibnamefont {Perc}}, \bibinfo {author} {\bibfnamefont {L.}~\bibnamefont
  {Shi}}, \ and\ \bibinfo {author} {\bibfnamefont {Z.}~\bibnamefont {Wang}},\
  }\href {\doibase 10.1007/s11071-022-07289-7} {\bibfield  {journal} {\bibinfo
  {journal} {Nonlinear Dynamics}\ }\textbf {\bibinfo {volume} {108}},\ \bibinfo
  {pages} {1837} (\bibinfo {year} {2022})}\BibitemShut {NoStop}%
\bibitem [{\citenamefont {Wang}\ \emph {et~al.}(2023)\citenamefont {Wang},
  \citenamefont {Fan}, \citenamefont {Zhang}, \citenamefont {Zou},\ and\
  \citenamefont {Wang}}]{Wang2023Synergistic}%
  \BibitemOpen
  \bibfield  {author} {\bibinfo {author} {\bibfnamefont {L.}~\bibnamefont
  {Wang}}, \bibinfo {author} {\bibfnamefont {L.}~\bibnamefont {Fan}}, \bibinfo
  {author} {\bibfnamefont {L.}~\bibnamefont {Zhang}}, \bibinfo {author}
  {\bibfnamefont {R.}~\bibnamefont {Zou}}, \ and\ \bibinfo {author}
  {\bibfnamefont {Z.}~\bibnamefont {Wang}},\ }\href {\doibase
  10.1088/1367-2630/acd26e} {\bibfield  {journal} {\bibinfo  {journal} {New
  Journal of Physics}\ }\textbf {\bibinfo {volume} {25}},\ \bibinfo {pages}
  {073008} (\bibinfo {year} {2023})}\BibitemShut {NoStop}%
\bibitem [{\citenamefont {Ding}\ \emph {et~al.}(2023)\citenamefont {Ding},
  \citenamefont {Zheng}, \citenamefont {Cai}, \citenamefont {Cai},
  \citenamefont {Chen}, \citenamefont {Zhang},\ and\ \citenamefont
  {Wang}}]{Ding2023emergence}%
  \BibitemOpen
  \bibfield  {author} {\bibinfo {author} {\bibfnamefont {Z.-W.}\ \bibnamefont
  {Ding}}, \bibinfo {author} {\bibfnamefont {G.-Z.}\ \bibnamefont {Zheng}},
  \bibinfo {author} {\bibfnamefont {C.-R.}\ \bibnamefont {Cai}}, \bibinfo
  {author} {\bibfnamefont {W.-R.}\ \bibnamefont {Cai}}, \bibinfo {author}
  {\bibfnamefont {L.}~\bibnamefont {Chen}}, \bibinfo {author} {\bibfnamefont
  {J.-Q.}\ \bibnamefont {Zhang}}, \ and\ \bibinfo {author} {\bibfnamefont
  {X.-M.}\ \bibnamefont {Wang}},\ }\href {\doibase
  https://doi.org/10.1016/j.chaos.2023.114032} {\bibfield  {journal} {\bibinfo
  {journal} {Chaos, Solitons \& Fractals}\ }\textbf {\bibinfo {volume} {175}},\
  \bibinfo {pages} {114032} (\bibinfo {year} {2023})}\BibitemShut {NoStop}%
\bibitem [{\citenamefont {Zheng}\ \emph
  {et~al.}(2023{\natexlab{a}})\citenamefont {Zheng}, \citenamefont {Zhang},
  \citenamefont {Zhang}, \citenamefont {Cai},\ and\ \citenamefont
  {Chen}}]{Zheng2023decoding}%
  \BibitemOpen
  \bibfield  {author} {\bibinfo {author} {\bibfnamefont {G.}~\bibnamefont
  {Zheng}}, \bibinfo {author} {\bibfnamefont {J.}~\bibnamefont {Zhang}},
  \bibinfo {author} {\bibfnamefont {J.}~\bibnamefont {Zhang}}, \bibinfo
  {author} {\bibfnamefont {W.}~\bibnamefont {Cai}}, \ and\ \bibinfo {author}
  {\bibfnamefont {L.}~\bibnamefont {Chen}},\ }\href@noop {} {\bibfield
  {journal} {\bibinfo  {journal} {arXiv preprint: 2309.14598}\ } (\bibinfo
  {year} {2023}{\natexlab{a}})}\BibitemShut {NoStop}%
\bibitem [{\citenamefont {Zhang}\ \emph {et~al.}(2019)\citenamefont {Zhang},
  \citenamefont {Dong}, \citenamefont {Liu}, \citenamefont {Huang},
  \citenamefont {Huang},\ and\ \citenamefont {Lai}}]{Zhang2019reinforcement}%
  \BibitemOpen
  \bibfield  {author} {\bibinfo {author} {\bibfnamefont {S.-P.}\ \bibnamefont
  {Zhang}}, \bibinfo {author} {\bibfnamefont {J.-Q.}\ \bibnamefont {Dong}},
  \bibinfo {author} {\bibfnamefont {L.}~\bibnamefont {Liu}}, \bibinfo {author}
  {\bibfnamefont {Z.-G.}\ \bibnamefont {Huang}}, \bibinfo {author}
  {\bibfnamefont {L.}~\bibnamefont {Huang}}, \ and\ \bibinfo {author}
  {\bibfnamefont {Y.-C.}\ \bibnamefont {Lai}},\ }\href {\doibase
  https://doi.org/10.1103/PhysRevE.99.032302} {\bibfield  {journal} {\bibinfo
  {journal} {Physical Review E}\ }\textbf {\bibinfo {volume} {99}},\ \bibinfo
  {pages} {032302} (\bibinfo {year} {2019})}\BibitemShut {NoStop}%
\bibitem [{\citenamefont {Zheng}\ \emph
  {et~al.}(2023{\natexlab{b}})\citenamefont {Zheng}, \citenamefont {Cai},
  \citenamefont {Qi}, \citenamefont {Zhang},\ and\ \citenamefont
  {Chen}}]{Zheng2023optimal}%
  \BibitemOpen
  \bibfield  {author} {\bibinfo {author} {\bibfnamefont {G.}~\bibnamefont
  {Zheng}}, \bibinfo {author} {\bibfnamefont {W.}~\bibnamefont {Cai}}, \bibinfo
  {author} {\bibfnamefont {G.}~\bibnamefont {Qi}}, \bibinfo {author}
  {\bibfnamefont {J.}~\bibnamefont {Zhang}}, \ and\ \bibinfo {author}
  {\bibfnamefont {L.}~\bibnamefont {Chen}},\ }\href@noop {} {\bibfield
  {journal} {\bibinfo  {journal} {arXiv preprint: 2312.14970}\ } (\bibinfo
  {year} {2023}{\natexlab{b}})}\BibitemShut {NoStop}%
\bibitem [{\citenamefont {Axelrod}\ and\ \citenamefont
  {Hamilton}(1981)}]{Robert1981Evolution}%
  \BibitemOpen
  \bibfield  {author} {\bibinfo {author} {\bibfnamefont {R.}~\bibnamefont
  {Axelrod}}\ and\ \bibinfo {author} {\bibfnamefont {W.~D.}\ \bibnamefont
  {Hamilton}},\ }\href {\doibase 10.1126/science.7466396} {\bibfield  {journal}
  {\bibinfo  {journal} {Science}\ }\textbf {\bibinfo {volume} {211}},\ \bibinfo
  {pages} {1390} (\bibinfo {year} {1981})}\BibitemShut {NoStop}%
\bibitem [{\citenamefont {Sutton}\ and\ \citenamefont
  {Barto}(2018)}]{Sutton2018reinforcement}%
  \BibitemOpen
  \bibfield  {author} {\bibinfo {author} {\bibfnamefont {R.~S.}\ \bibnamefont
  {Sutton}}\ and\ \bibinfo {author} {\bibfnamefont {A.~G.}\ \bibnamefont
  {Barto}},\ }\href {https://ieeexplore.ieee.org/servlet/opac?bknumber=6267343}
  {\emph {\bibinfo {title} {Reinforcement Learning: An Introduction}}}\
  (\bibinfo  {publisher} {MIT press},\ \bibinfo {year} {2018})\BibitemShut
  {NoStop}%
\bibitem [{\citenamefont {Volpe}(2021)}]{Sethna2021statistical}%
  \BibitemOpen
  \bibfield  {author} {\bibinfo {author} {\bibfnamefont {G.}~\bibnamefont
  {Volpe}},\ }\href {\doibase 10.1080/00107514.2021.2002948} {\bibfield
  {journal} {\bibinfo  {journal} {Contemporary Physics}\ }\textbf {\bibinfo
  {volume} {62}},\ \bibinfo {pages} {121} (\bibinfo {year} {2021})}\BibitemShut
  {NoStop}%
\bibitem [{\citenamefont {Cai}\ \emph {et~al.}(2015)\citenamefont {Cai},
  \citenamefont {Chen}, \citenamefont {Ghanbarnejad},\ and\ \citenamefont
  {Grassberger}}]{Cai2015avalanche}%
  \BibitemOpen
  \bibfield  {author} {\bibinfo {author} {\bibfnamefont {W.}~\bibnamefont
  {Cai}}, \bibinfo {author} {\bibfnamefont {L.}~\bibnamefont {Chen}}, \bibinfo
  {author} {\bibfnamefont {F.}~\bibnamefont {Ghanbarnejad}}, \ and\ \bibinfo
  {author} {\bibfnamefont {P.}~\bibnamefont {Grassberger}},\ }\href {\doibase
  https://doi.org/10.1038/nphys3457} {\bibfield  {journal} {\bibinfo  {journal}
  {Nature Physics}\ }\textbf {\bibinfo {volume} {11}},\ \bibinfo {pages} {936}
  (\bibinfo {year} {2015})}\BibitemShut {NoStop}%
\bibitem [{\citenamefont {Huang}(2015)}]{Huang2015Experimental}%
  \BibitemOpen
  \bibfield  {author} {\bibinfo {author} {\bibfnamefont {J.}~\bibnamefont
  {Huang}},\ }\href {\doibase https://doi.org/10.1016/j.physrep.2014.11.005}
  {\bibfield  {journal} {\bibinfo  {journal} {Physics Reports}\ }\textbf
  {\bibinfo {volume} {564}},\ \bibinfo {pages} {1} (\bibinfo {year}
  {2015})}\BibitemShut {NoStop}%
\end{thebibliography}%
\end{document}